\documentclass[a4paper,11pt]{article}
\pdfoutput=1
\usepackage{jheppub}
\usepackage[T1]{fontenc} 
\usepackage{fixltx2e}

\usepackage{pdflscape}
\usepackage{afterpage}
\usepackage{vmargin,amssymb,amsmath,epsfig,latexsym,booktabs,braket}

\newcommand{\tr}{\text{Tr}}
\newcommand{\T}{\mathbf{T}}
\renewcommand{\P}{\mathbf{P}}
\newcommand{\R}{\mathbf{R}}
\newcommand{\X}{\mathcal{X}}
\newcommand{\Y}{\mathcal{Y}}
\newcommand{\Z}{\mathcal{Z}}

\newcommand{\la}{\langle}
\newcommand{\ra}{\rangle}
\newcommand{\cN}{{\cal N}}
\newcommand{\cA}{{\cal A}}
\newcommand{\cB}{{\cal B}}
\newcommand{\cG}{{\cal G}}
\newcommand{\cO}{{\cal O}}
\newcommand{\mh}{\mathfrak{h}}
\newcommand{\hhb}[6]{\widehat{\mathfrak{h}}_{#1 #2 #3}(\{#4\},\emptyset,\{#6\})}
\newcommand{\hhc}[6]{\widehat{\mathfrak{h}}_{#1 #2 #3}(\{#4\},\{#5\},\emptyset)}
\newcommand{\pp}[1]{\frac{y^2_{#1}}{x^2_{#1}}}

\newcommand{\Af}{\stackrel{\rightarrow}{A}}
\newcommand{\Ab}{\stackrel{\leftarrow}{A}}
\newcommand{\Bf}{\stackrel{\rightarrow}{B}}
\newcommand{\Bb}{\stackrel{\leftarrow}{B}}

\let\emptyset\varnothing

\title{\boldmath Colour-dressed hexagon tessellations for correlation functions and non-planar corrections}

\preprint{HU-EP-17/23, HU-MATH 2017-07}

\author[a]{Burkhard Eden,}
\author[b]{Yunfeng Jiang,}
\author[a]{Dennis le Plat,}
\author[b]{Alessandro Sfondrini}

\affiliation[a]{Institut f\"{u}r Mathematik und Physik, Humboldt-Universit{\"a}t zu Berlin, \\
Zum gro{\ss}en Windkanal 6, 12489 Berlin, Germany}
\affiliation[b]{Institut f\"ur theoretische Physik, ETH Z\"urich, \\
Wolfgang-Pauli-Stra{\ss}e 27, 8093 Z\"urich, Switzerland}

\emailAdd{eden@math.hu-berlin.de}
\emailAdd{jiangyu@phys.ethz.ch}
\emailAdd{diplat@physik.hu-berlin.de}
\emailAdd{sfondria@itp.phys.ethz.ch}

\abstract{We continue the study of four-point correlation functions by the \textit{hexagon tessellation} approach initiated in arXiv:\href{https://arxiv.org/abs/1611.05436}{1611.05436} and arXiv:\href{https://arxiv.org/abs/1611.05577}{1611.05577}.
We consider planar tree-level correlation functions in $\cN = 4$ supersymmetric Yang-Mills theory involving two non-protected operators.
We find that, in order to reproduce the field theory result, it is necessary to include $SU(N)$ colour factors in the hexagon formalism; moreover, we find that the hexagon approach as it stands is naturally tailored to the single-trace part of correlation functions, and does not account for multi-trace admixtures. We discuss how to compute correlators involving double-trace operators, as well as more general $1/N$ effects; in particular we compute the whole next-to-leading order in the large-$N$ expansion of tree-level BMN two-point functions by tessellating a torus with punctures.
Finally, we turn to the issue of ``wrapping'', L\"uscher-like corrections. We show that $SU(N)$ colour-dressing reproduces an earlier empirical rule for incorporating single-magnon wrapping, and we provide a direct interpretation of such wrapping processes in terms of $\mathcal{N}=2$ supersymmetric Feynman diagrams.}

\begin{document}
\maketitle
\flushbottom

\section*{}
\thispagestyle{empty}

\newpage
\section{Introduction}
The first objects to be studied in the framework of the AdS/CFT correspondence \cite{Maldacena:1997re,Gubser:1998bc,Witten:1998qj} were correlation functions of BPS operators in $\cN = 4$ supersymmetric Yang-Mills theory ($\cN = 4$ SYM), i.e. gauge-invariant composite operators without anomalous dimensions. In particular, non-renormalised correlation functions received a great deal of attention, see \textit{e.g.}~\cite{Freedman:1998tz,Lee:1998bxa,Eden:1999gh}, because they  have to coincide in weak-coupling perturbation theory and in supergravity.
What is more interesting, though, are quantities that do receive quantum corrections. The study of the spectrum of anomalous dimensions of the ``BMN'' operators~\cite{Berenstein:2002jq} in the large-$N$ limit~\cite{'tHooft:1973alw} and of the energy levels of the dual string states was the starting point for an astounding development: the spectral problem was mapped to the solution of an integrable spin chain \cite{Minahan:2002ve} at weak 't~Hooft coupling. This has been extended to all composite operators of the theory and to arbitrarily high loop order~\cite{Beisert:2004hm,Arutyunov:2004vx, Beisert:2005fw,Beisert:2006ez}. The picture was then completed by incorporating the so-called ``wrapping'' finite-size effects~\cite{Ambjorn:2005wa}, which could be done from the point of view of the dual string theory, see refs.~\cite{Arutyunov:2009ga, Beisert:2010jr,Gromov:2017blm} for reviews;
this allows the computation of the spectrum up to amazingly high orders in the perturbative expansion in the 't~Hooft coupling, or numerically at finite coupling with great precision.
The discussion of the spectrum of anomalous dimensions of the $\cN = 4$ SYM along these lines is thus by now complete, at least in principle.

The study of three-point functions of non-protected composite operators by integrability was initiated only much later in ref.~\cite{Escobedo:2010xs}. Substantial progress came about very recently by the introduction of the \emph{hexagon form-factor approach}~\cite{Basso:2015zoa}. The key to this approach is to consider the string worldsheet with three punctures, and cut it into two hexagonal patches. Each of these is interpreted as containing a non-local operator that creates a conical excess---the hexagon operator. The ``asymptotic'' three point function is reconstructed by summing over the form factors of such an operator; this is the part that discards wrapping effects. These too can be included in the formalism~\cite{Basso:2015zoa,Eden:2015ija, Basso:2015eqa,Basso:2017muf}, even though at the current stage it is possible to do so only in a magnon-by-magnon manner reminiscent of L\"uscher corrections. It would be desirable to have a TBA-like approach which takes into account all the wrapping corrections at once. In parallel, integrability for the string-field theory vertex has been explored, too~\cite{Jiang:2014cya, Jiang:2014gha, Bajnok:2015hla,Bajnok:2017mdf}.

\begin{figure}[t]
\begin{center}
\includegraphics[width=0.7\linewidth]{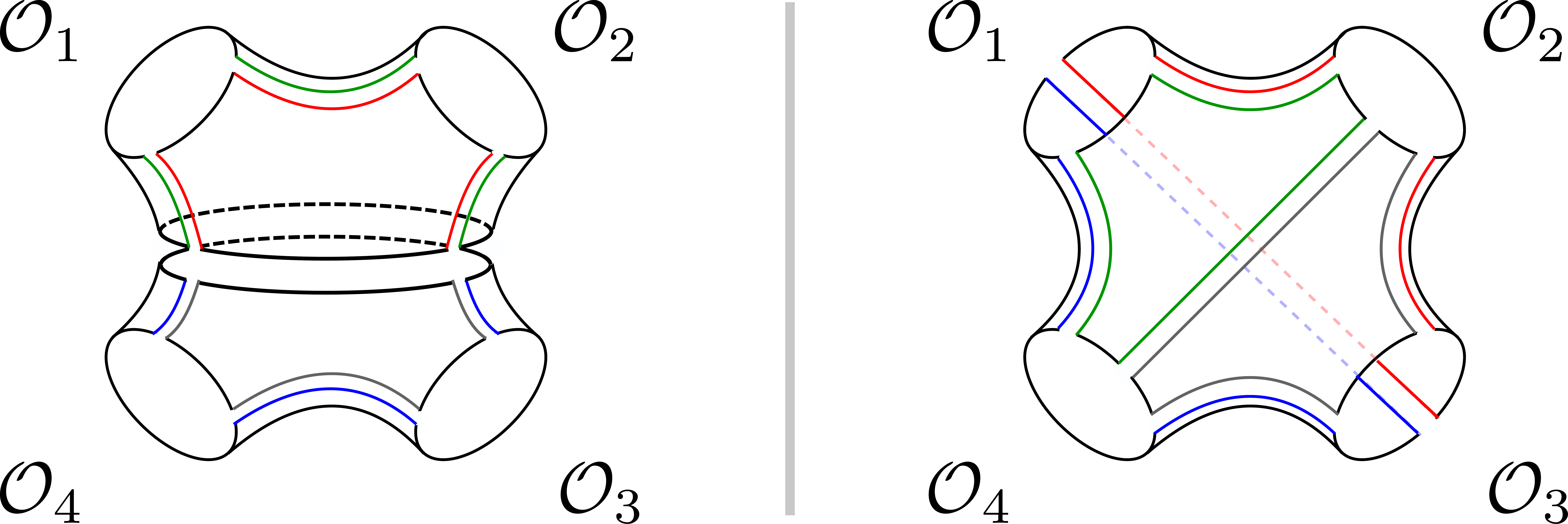}
\caption{We depict two ways of tessellating a four-point function by hexagonal patches. On the left, we first cut it into two three-point functions, which can then be cut into hexagons; this corresponds to performing the operator product expansion (OPE), and it requires summing over intermediate physical states. On the right, we cut the four point-functions into four hexagons without introducing a sum over intermediate physical states.
}\label{fig:4pttwocuts}
\end{center}
\end{figure}

A natural next step is the study of four-point functions. These are rich objects, since unlike lower-point functions they have a non-trivial (and intricate) dependence on the position of the operators through the conformal cross-ratios. As they encode information on lower-point functions, they also play a crucial role in the conformal bootstrap approach, see \textit{e.g.}\ ref.~\cite{Simmons-Duffin:2016gjk}. In AdS/CFT, their study was undertaken early on, focussing especially on $\tfrac{1}{2}$-BPS operators, see \textit{e.g.}\ refs.~\cite{Arutyunov:1999fb,DHoker:1999kzh,Eden:2000mv,Bianchi:2000hn}. More recently, it was understood that four-point correlators capture information on locality in the bulk~\cite{Heemskerk:2009pn,Maldacena:2015iua}, which makes their investigation particularly interesting. Indeed in recent times a number of new results have appeared in this field~\cite{Rastelli:2016nze,Arutyunov:2017dti, Rastelli:2017udc}.
In principle, four-point functions are fixed via the operator product expansion (OPE) in terms of the lower-point correlators. In practice, resumming the OPE is a daunting tasks, and would require accounting for multi-trace operators---a difficult task in integrability, so far.%
\footnote{The OPE approach has been recently considered in the context of integrability in ref.~\cite{Basso:2017khq}.}
 An alternative approach was advocated by two of us in ref.~\cite{Eden:2016xvg} and independently in ref.~\cite{Fleury:2016ykk} by Fleury and Komatsu, building on the hexagon proposal~\cite{Basso:2015zoa} and on earlier investigation of four-point functions in integrability~\cite{Caetano:2011eb}. The idea is to tessellate the four-point function by hexagons, without cutting it into two three-point functions like in the OPE,%
\footnote{This tessellation approach is reminiscent of the ``pentagon'' approach for scattering amplitudes~\cite{Alday:2010ku,Basso:2013vsa}.}
see figure~\ref{fig:4pttwocuts}.
Moreover, it is necessary to  include the dependence on the conformal cross-ratios in the hexagon approach. The computation of the asymptotic part of the four-point function is then straightforward. Including wrapping effects is also possible, as it was discussed for a single ``mirror magnon'' in ref.~\cite{Fleury:2016ykk}, though this requires some empirical rules on which diagrams to include in the computation of wrapping corrections.

In this paper, we continue the investigation of hexagon tessellations. While in refs.~\cite{Eden:2016xvg,Fleury:2016ykk} four-point functions involving at most \textit{one} non-protected operator were considered, here we increase the complexity of the set-up and allow for \textit{two} non-protected operators. This might seem a slight technical complication; yet there are important conceptual lessons to be learned, even at tree-level.

Firstly, we see that certain connected, but one-particle reducible diagrams have to be excluded from tree-level hexagon tessellations---something that had not been anticipated from earlier studies. We propose that the correct way to account for this is to include $SU(N)$ colour factors in the hexagon formalism. Not only this prescription allows us to non-trivially reproduce several field-theory results, but it also automatically incorporates the empirical rules for wrapping at one-loop proposed in ref.~\cite{Fleury:2016ykk}.

Secondly, we see that the hexagon formalism does not capture multi-trace admixtures, even when those give leading effects in the $1/N$ expansion in field theory. This is not entirely surprising, given that the whole integrability approach is naturally tailored to single-trace operators. However it does raise the question of how to include such effects.
Indeed it is an outstanding challenge to account for multi-trace operators and, more in general, $1/N$ effects in the integrability, see \textit{e.g.}\ ref.~\cite{Kristjansen:2010kg}. There are three facets to this problem; to begin with, correlation functions should not only be represented on a sphere, but also on higher-genus surfaces; next, in general it is necessary to compute correlation functions that involve one or more multi-trace operators; finally, the precise structure of the mixing between single and multi-trace operators should be found by diagonalising the complete dilatation operator. In our work we focus on the computation of correlation functions, hence on the first two tasks.
In particular, we study two-point functions of non-protected operators at next-to-leading order in the $1/N$ expansion by hexagon tessellations. Indeed we find two contributions: from the higher genus topology (a torus at this order), and from single-trace--double-trace correlators. The first term can be studied by tessellating a torus by hexagons%
\footnote{The idea of tessellating higher-genus surfaces by hexagons was also proposed by P.~Vieira~\cite{vieira:talk}.}.
In practice, we reduce the problem to computing a four-point function on a torus reminiscent of what was considered in ref.~\cite{Beisert:2002bb}; to reproduce the two-point function we take two of these operators to be the identity, as it is done when computing the Gaudin norm by hexagons~\cite{Basso:2015zoa}. Also in this context, colour-dressing turns out to be essential to reproduce the field-theory result. Correlation functions involving double-trace operators can also be dealt with by similar identity insertions.

Finally, we turn to wrapping corrections, showing that indeed colour-dressing gives the correct rules for selecting which diagrams to dress by mirror magnons, at least in the one-loop case which is the only one studied in the literature so far. In the process, we find a direct relation between single-magnon exchanges in the hexagon formalism and Yang-Mills lines in the $\cN=2$ supersymmetric Feynman diagram formalism.

The paper is organised as follows: in section~\ref{sec:treelvl}, we review the computation of four-point functions at tree level in field theory for the case at hand; we also introduce the Drukker-Plefka restricted kinematics~\cite{Drukker:2009sf,Bargheer:2017eoz}, which is natural for the hexagon formalism~\cite{Basso:2015zoa} and which we will employ for several computations throughout the paper. In section~\ref{sec:hexagon} we briefly review the hexagon formalism starting from the case of three-point functions; we discuss the case of four-point functions at some length and comment on how the approach of ref.~\cite{Eden:2016xvg} relates to the one of ref.~\cite{Fleury:2016ykk}. In section~\ref{sec:twoBMN} we detail the computation of planar four-point functions over two protected and two non-protected operators, explain the need for $SU(N)$ colour dressing and introduce the study of double-trace admixtures. In section~\ref{sec:1overN} we show how to use the hexagon formalism to compute the next-to-leading-order in the $1/N$ expansion for the tree-level two-point function of non-protected operators. In section~\ref{sec:Nis2} we show that colour-dressing automatically encodes the known rules for wrapping processes at one loop, and propose an interpretation of wrapping modes in terms of $\cN=2$ supersymmetric Feynman diagrams. We conclude in section~\ref{sec:conclusions}, and relegate some details to the appendices.

\section{Tree-level four-point functions with two non-protected operators} \label{sec:treelvl}
We consider four-point functions of scalar operators in $\mathcal{N}=4$ supersymmetric Yang-Mills theory. The simplest operators that we can consider are $\tfrac{1}{2}$-BPS ones, such as $\tr[Z^L]$ where $Z$ is a complex scalar. Starting from such an operator it is possible to define \textit{supersymmetry-protected} four-point functions, by considering $\tfrac{1}{2}$-BPS operators in a special kinematic configuration~\cite{Drukker:2009sf}. We introduce the super-translation $\T$,
\begin{equation}
\T=- i \epsilon_{\alpha\dot{\alpha}}\P^{\alpha\dot{\alpha}} +\epsilon_{a\dot{a}}\R^{a\dot{a}},
\end{equation}
written in terms of the Poincar\'e translation $\P^{\alpha\dot{\alpha}}$ and of the R-symmetry generator~$\R^{a\dot{a}}$. From a scalar $Z$ at position $x^\mu=0$ we hence have
\begin{equation}
\label{eq:drukkerplefka}
\Z(a):=e^{a\T}Z(0)e^{-a\T}=[Z+a(Y-\bar{Y})+a^2\bar{Z}](0,a,0,0)\,.
\end{equation}
Similarly, we find
\begin{equation}
\Y(a)=[Y+a\bar{Z}](0,a,0,0),\qquad
\bar{\Y}(a)=[\bar{Y}-a\bar{Z}](0,a,0,0),
\end{equation}
while $X$ and $\bar{X}$ are only translated in Minkowski space. We will suppress the $a$-dependence when this does not cause confusion, or adopt the short-hand notation $\Z_j=\Z(a_j)$. It is useful to spell out the non-vanishing propagators
\begin{equation}
\label{eq:propagators}
\la \X_i \bar{\X}_j\ra=\frac{1}{a_{ij}^2},\qquad
\la \Y_i \bar{\Y}_j\ra=\frac{1}{a_{ij}^2},\qquad
\la \Y_i {\Z}_j\ra=\frac{1}{a_{ij}},\qquad
\la \bar{\Y}_i {\Z}_j\ra=\frac{1}{a_{ji}},\qquad
\la \Z_i {\Z}_j\ra=1\,,
\end{equation}
where $a_{ij}=-a_{ji}=a_i-a_j$. We denote $\tfrac{1}{2}$-BPS operators of length~$L$ as
\begin{equation}
\cO_L = \frac{1}{\sqrt{L N^L}} \tr(\Z^L).
\end{equation}
Computing the tree-level four-point functions of such operators is straightforward, though somewhat cumbersome, and can be done by taking Wick contractions and using eq.~\eqref{eq:propagators}. Moreover, these correlators are protected by supersymmetry so that the tree-level result does not get corrected at higher loops~\cite{Drukker:2009sf}.

One way to obtain more interesting correlators is to allow some of the operators to be non-protected. We shall focus on so-called BMN operators~\cite{Berenstein:2002jq} with two impurities. It is only a technical complication to consider more general operators, both in field theory and in the hexagon approach of the next section. For the purposes of this paper, we will focus on this simplest non-trivial example.
Hence we consider operators of the type
\begin{equation}
\label{eq:bmn}
\cO_L^k = \tr(\Z^{L-k-2} \Y \Z^k \Y) \, ,
\end{equation}
where we could (and will) also allow the two impurities to be $\{\bar \Y, \bar \Y\}$, $\{\X,\X\}$ or $\{\bar \X, \bar \X\}$. For each set of excitations, eq.~\eqref{eq:bmn} gives $\lfloor L/2\rfloor$ distinct operators. Conformal eigenstates are found by diagonalising the dilatation operator for every $L$, which gives a linear combination of~$\cO_L^k$s with definite anomalous dimension~$\gamma=g^2\gamma_1+O(g^4)$..
We are interested in operators with $\gamma_1\neq0$, as these are independent from the $\tfrac{1}{2}$-BPS states---rather than being a symmetry descendant thereof.

\begin{table}[t]
\begin{center}
\begin{tabular}{l|l|l|l|c}
$L$ & operator & Bethe vector & $\gamma_1$ & Rapidity $u$ \\[1 mm]
\hline
4 & $\cB_4$ & $\frac{1}{\sqrt{3 N^4}}\big(\cO_4^0 - \cO_4^1\big)$ & 6 & $\frac{1}{2 \sqrt{3}}$ \\[2 mm]
5 & $\cB_5$ & $\frac{1}{\sqrt{2 N^5}}\big(\cO_5^0 - \cO_5^1\big)$ & 4 & $\frac{1}{2}$ \\[2 mm]
6 & $\cB_6^\mp$ & $\frac{1}{\sqrt{5 N^6}}\big(\frac{1 \pm \sqrt{5}}{2} \cO_6^0 + \frac{1 \mp \sqrt{5}}{2} \cO_6^1 - \cO_6^2\big)$ & $5 \mp \sqrt{5}$ & $\frac{1}{2} \sqrt{ 1 \pm \frac{2}{\sqrt{5}}}$ \\[2 mm]
7 & $\cB_7'$ & $\frac{1}{\sqrt{2 N^7}}\big(\cO_7^0 - \cO_7^2\big)$ & 2 & $\frac{\sqrt{3}}{2}$ \\[2 mm]
7& $\cB_7''$ & $\frac{1}{\sqrt{6 N^7}}\big(\cO_7^0 - 2 \cO_7^1 + \cO_7^2\big)$ & 6 & $\frac{1}{2 \sqrt{3}}$
\end{tabular}
\caption{BMN operators with two impurities and length $L\leq 7$. we also list the one-loop anomalous dimension $\gamma_1$ and the Bethe rapidity $u=u_1$, \textit{cf.}\ eq.~\eqref{eq:betheBMN}. Operators are normalised so that their two-point function reads $\langle\cB_L\cB_L\rangle=1+O(1/N^2)$.}
\label{tab:adim}
\end{center}
\end{table}

It is a well-known yet remarkable fact that such operators and their planar anomalous dimensions can be found by computing the spectrum of an integrable $SU(2)$ spin chain with nearest-neighbour interactions~\cite{Minahan:2002ve}. The Bethe ansatz equations for a chain of length $L$ take a simple form in terms of the ``rapidities'' $u_1,u_2$ of the two impurities; these are associated to the momentum as
\begin{equation}
e^{i \, p(u)} = \frac{u+i/2}{u-i/2}\,,
\end{equation}
in our convention. Furthermore, cyclicity of the trace requires $u_1+u_2=0$ so that the Bethe ansatz equation and the anomalous dimension are simply given by~\cite{Minahan:2002ve}
\begin{equation}
\left(\frac{u_1+i/2}{u_1 - i/2}\right)^{L-1}=1 \, ,\qquad
u_2=-u_1\,,
\qquad \gamma_1 = \sum_{i=1}^2 \frac{1}{u_i^2 + 1/4} \, . \label{eq:betheBMN}
\end{equation}
The eigenvectors of the dilatation operator, in the planar limit, are given by the Bethe wave-functions associated to a given rapidity. In table~\ref{tab:adim} we list the first few eigenstates with $\gamma_1\neq0$, along with their associated anomalous dimension and rapidity. Notice that the eigenstates have been normalised for later convenience.

From this set of operators, it is also straightforward to compute tree-level correlation functions. Our focus here is on four-point functions involving two $\tfrac{1}{2}$-BPS operators and two (non-protected) BMN operators. We will list several such correlators in table~\ref{tab:fieldtheroycorr} below; in the next section we will see how to reproduce that table using the integrability-based approach of hexagon tessellations. Finally we would like to mention that we restrict ourselves to the simplest BMN operators with two magnons in order to keep the discussions as simple as possible. We can, of course, perform the same analysis for longer operators with more magnons. This would increase the complexity of our exercise without revealing any further insight.

\section{The hexagon formalism for correlation functions}
\label{sec:hexagon}
We start by reviewing the hexagon approach to correlation functions.

\subsection{The hexagon proposal for three-point functions}
Above we have discussed correlation functions in $\cN=4$ SYM. In the dual string theory, $n$-point correlation functions emerge from puncturing the string worldsheet $n$~times. The simplest case, for $n=3$, gives the topology of a ``pair of pants''. It was suggested in ref.~\cite{Basso:2015zoa} that the three-point function can be found from decompactifying the worldsheet  by cutting the pants ``along the seams'', which gives two hexagonal patches. Each of these hexagonal patches could be mapped to an ordinary (square) worldsheet patch with the insertion of a conical excess operator---the \textit{hexagon} operator. A sum over form factors of this operator will then yield the three-point function. This generalises the cutting and sewing of spin~chains that is natural at weak 't~Hooft coupling~\cite{Escobedo:2010xs} with the advantage that the hexagon form factor is known non-perturbatively---much like what happens for the S~matrix for two point functions~\cite{Beisert:2005tm,Arutyunov:2006yd}.

Let us briefly illustrate the construction on a simple case: a three-point function of the form $\langle  \cB_{L_1} \cO_{L_2} \cO_{L_3}\rangle$, involving one non-protected operator and two $\tfrac{1}{2}$-BPS operators.
We start from three operators $\cO_{L_1},\cO_{L_2},\cO_{L_3}$ which without loss of generality we can take at positions $a_1=0$, $a_2=1$ and $a_3=\infty$. Decompactifying  the resulting three-point function gives two empty hexagons; to get the $\langle \cB_{L_1} \cO_{L_2} \cO_{L_3}\rangle$ three-point function we have to introduce two impurities (magnons) with rapidities $u_1,u_2$ on top of the vacuum $\cO_{L_1}$. Then, the hexagon tessellation yields a sum over partitions $\alpha,\bar{\alpha}$ with $\alpha\cup\bar{\alpha}=\{u_1,u_2\}$. The sum is weighted by phases that emerge from transporting either magnon across the chain~\cite{Basso:2015zoa},
\begin{equation}
\label{eq:oldmathcalA}
\mathcal{A}=\sum_{\bf{u}=\alpha\cup\bar{\alpha}} (-1)^{|\bar{\alpha}|}\, \omega(\alpha,\bar{\alpha},\ell_{12}) \,\mathfrak{h}_{\text{123}}(\alpha)\,\mathfrak{h}_{\text{132}}(\bar{\alpha}),\qquad
\omega(\alpha,\bar{\alpha},\ell)=\prod_{k\in\bar{\alpha}} e^{ip_k\ell}\prod_{j\in\alpha, j>k}S_{kj},
\end{equation}
In particular, for a two-impurity state we have \textit{e.g.}
\begin{equation}
\begin{aligned}
\mathcal{A}= \mathfrak{h}_{\text{123}}(\{u_1,u_2\})\,\mathfrak{h}_{\text{132}}(\emptyset)
-e^{ip_2\ell_{12}} \mathfrak{h}_{\text{123}}(\{u_1\})\,\mathfrak{h}_{\text{132}}(\{u_2\})\qquad\qquad\qquad\qquad\\
-S_{12}e^{ip_1\ell_{12}} \mathfrak{h}_{\text{123}}(\{u_2\})\,\mathfrak{h}_{\text{132}}(\{u_1\})
+e^{i (p_1+p_2)\ell_{12}}\mathfrak{h}_{\text{123}}(\emptyset)\,\mathfrak{h}_{\text{132}}(\{u_1,u_2\})\,,
\end{aligned}
\end{equation}
where $\emptyset$ is the empty set.
The ingredients in this formula are Beisert's S-matrix elements~\cite{Beisert:2005tm} for the scattering of the two impurities, and the hexagon form factor, which also depends on the impurities we consider. The empty hexagon is normalised to give $\mathfrak{h}(\emptyset)=1$; in appendix~\ref{app:tree} we collect the tree-level expressions needed for the computation of the hexagon form factors with flavours~$Z, Y,\bar{Y} ,X,\bar{X}$. From~$\mathcal{A}$, the three-point function follows immediately as~\cite{Basso:2015zoa}
\begin{equation}
\label{eq:BKVformula}
\langle \cB_{L_1}^{k}\cO_{L_2}\cO_{L_3}\rangle
=
\frac{\sqrt{L_1L_2L_3}}{N}
\frac{\mathcal{A}}{\sqrt{\mathcal{G}}\prod_{j<k}\sqrt{S_{jk}}},
\end{equation}
where the Gaudin norm~$\mathcal{G}$ is given by
\begin{equation}
\mathcal{G}=\text{det}\left[\frac{\partial \Phi_j(\mathbf{u})}{\partial u_k}\right]\quad \text{with}\quad
e^{i \Phi_{j}(\mathbf{u})}=e^{ip_j\ell}\prod_{i\neq j}^2 S(u_j,u_i)\,.
\end{equation}
Notice that this construction is only asymptotic, and it should be completed by incorporating L\"uscher-like finite-size corrections~\cite{Basso:2015zoa,Eden:2015ija, Basso:2015eqa,Basso:2017muf}. However, we will not need the details of wrapping corrections in the rest of this paper, and we will not review them here.

\subsection{Four-point functions and position-dependence for hexagons}
\label{sec:positiondressing}
It is not immediately obvious how to adapt the above construction to describe \textit{four-point} correlation functions. Perhaps the most glaring issue is that only \textit{three} points can be put at chosen locations---say $0,1,\infty$---and the final, physical result will depend on the position of the fourth point. Even in the restricted kinematics~\cite{Drukker:2009sf} described in section~\ref{sec:treelvl}, this introduces a new parameter~$a\in\mathbb{R}$. In ref.~\cite{Eden:2016xvg} we have proposed how to account for the dependence on the position and how to tessellate the four-point function. We will review that proposal in some detail, commenting on how it relates to the similar approach proposed independently in ref.~\cite{Fleury:2016ykk}.

A first indication of how to include position dependence (or, equivalently, R-symmetry charge dependence) comes from field theory, and was put forward in ref.~\cite{Eden:2016xvg}. Let us start from the simplest case of a hexagon involving three operators, two of which are $\tfrac{1}{2}$-BPS. The remaining operator, which we take at point $(x_1)^\mu$, contains a single impurity---\textit{e.g.}\ arising from acting with a derivative $\partial^\mu$ in Lorenz space.  The resulting three-point function is a Lorenz-covariant tensor; besides, by conformal invariance, it must have definite conformal weight at point $(x_1)^\mu$. The only possibility is hence the conformal vector
\begin{equation}
\label{eq:vdef1}
(v_{1;23})^{\mu}=\frac{(x_{12})^\mu}{(x_{12})^\nu(x_{12})_\nu}-\frac{(x_{13})^\mu}{(x_{13})^\nu(x_{13})_\nu}\,.
\end{equation}
It is convenient to parametrise this vector by introducing the holomorphic and anti-holomorphic part%
\footnote{%
It is thanks to the four-point function kinematics that we can write our results in terms of holomorphic and anti-holomorphic coordinates. Putting three of the operators on a line in Minkowski space, which can be done without loss of generality owing to conformal invariance, the whole four-point function is defined on a plane, hence the two-dimensional kinematics; the same happens in R-symmetry space.
}
 of the distances $x_{ij}$, $x_{ij}^+$ and $x_{ij}^-$, respectively. This will allow us to make contact with ref.~\cite{Fleury:2016ykk}. In fact, projecting on the (anti-)holomorphic part and using $x^\mu x_\mu=x^+x^-$, we have
\begin{equation}
(v_{1;23})^{\pm}=\frac{x_{12}^\pm}{x_{12}^+x_{12}^-}-\frac{x_{13}^\pm}{x_{13}^+x_{13}^-}=\frac{x^{\mp}_{23}}{x^{\mp}_{12}x^{\mp}_{13}}\,.
\end{equation}
This is precisely the position-dressing associated to the hexagon in ref.~\cite{Fleury:2016ykk}, see also appendix~\ref{app:position}, even though the authors there reached this expression by a somewhat different reasoning.
Similarly, if instead of acting with a derivative $\partial^\mu$ we acted with a lowering operator $J^i$ in $su(4)$ R-symmetry space, we would have found a vector $(u_{1;23})^{i}$ in the embedding formalism. Again, this is most simply expressed in terms of holomorphic and anti-holomorphic parameters which we denote as $y_{ij}^\pm$.
 By the very same algebra we have
\begin{equation}
(u_{1;23})^{\pm}=\frac{y^{\mp}_{23}}{y^{\mp}_{12}y^{\mp}_{13}}\,.
\end{equation}
We refer the reader to appendix~\ref{app:position} for further details on this prescription and for its comparison with the approach of ref.~\cite{Fleury:2016ykk}. For the purpose of this paper we shall mostly restrict to the Drukker-Plefka kinematics. Notice that in eq.~\eqref{eq:drukkerplefka} we have coupled the Poincar\'e translation and the R-symmetry rotation; hence, the Minkowski and R-symmetry vectors are now related and in fact, for our choice, identical.%
\footnote{%
Running a bit ahead of ourselves, let us remark that in terms of the conformal cross-ratios the line configuration reads simply $z=\bar{z}=\alpha=\bar{\alpha}=a$, \textit{cf.}\  appendix~\ref{app:position}.
}
 Both vectors can be written in terms of the positions $a_j$; we find, for the only non-vanishing component of $(v_{1;23})^{\mu}$
\begin{equation}
\label{eq:vdef2}
v_{1;23}=u_{1;23}=\frac{1}{a_{12}}-\frac{1}{a_{13}}=\frac{a_{23}}{a_{12}a_{13}}\,.
\end{equation}
This argument can be repeated for excitations at any of the operators in a given hexagon, and it gives a simple prescription for incorporating space-time dependence at tree level; schematically~\cite{Eden:2016xvg}
\begin{equation}
\label{eq:hexagoncorr}
\begin{aligned}
\mathfrak{h}_{123}(\alpha_1,\alpha_2,\alpha_3)\quad \to\quad&
\widehat{\mathfrak{h}}_{123}(\alpha_1,\alpha_2,\alpha_3)\\
&=
v_{1;23}^{|\alpha_1|}\,v_{2;31}^{|\alpha_2|}\,v_{3;12}^{|\alpha_3|}\
\mathfrak{h}_{123}(\alpha_1,\alpha_2,\alpha_3)\,,
\end{aligned}
\end{equation}
where we have three groups of excitations, with $|\alpha_j|$ excitations at position $(x_j)^\mu=(0,a_j,0,0)$, for $j=1,\dots3$.
Beyond tree-level, one should take into account that the scaling dimension of magnons is corrected as in eq.~\eqref{eq:betheBMN} so that the exponents $|\alpha_j|$ are shifted by $\gamma(\alpha_i)$, the anomalous dimension of the magnons in the set $\alpha_i$~\cite{Fleury:2016ykk}. While we mostly work in the restricted kinematics on the line, it is rather straightforward to promote the vectors $v_{i;jk}$ to functions of the holomorphic coordinates, see appendix~\ref{app:position}.

We remark that $v_{i;jk}$s are clearly not independent for different choices of $i,j,k$. For instance, $v_{i;jk}+v_{i;kl}=v_{i;jl}$. In ref.~\cite{Eden:2016xvg} we found that the  tree-level four-point function of the form $\langle \cB_{L_1} \cO_{L_2}\cO_{L_3}\cO_{L_4}\rangle$ can be written in terms of the following basis
\begin{equation}
\big(v_{1;23}^2,\ v_{1;23}\,v_{1;24},\ v_{1;24}^2\big)\,.
\end{equation}
For the four-point functions involving \textit{two} BMN operators of table~\ref{tab:fieldtheroycorr}, we might na\"ively expect nine such basis elements; however, only five of those are linearly independent and we can therefore introduce the basis vector $\mathbf{v}$
\begin{equation}
\label{eq:basis}
\mathbf{v}=\big(
v^2_{1;24}\,v^2_{2;13},\
v_{1;23}\,v_{1;24}\,v^2_{2;13},\
v_{1;23}^2\,v_{2;13}^2,\
v_{1;23}^2\,v_{2;13}\,v_{2;14},\
v_{1;23}^2\,v_{2;14}^2
\big)\,.
\end{equation}
Using this, we can compactly write down the tree-level four-point functions as
\begin{equation}
\label{eq:4pbasis}
\langle \cB_{L_1}\,\cB_{L_2}\, \cO_{L_3}\,\cO_{L_4}\rangle
= \frac{1}{N^2} \, \mathbf{m}_{\chi,\chi'} \cdot \mathbf{v},
\end{equation}
where $\mathbf{m}_{\chi,\chi'}$ is a vector of coefficients depending on the impurities' flavours $\chi,\chi'$ which we may take to be $X,\bar{X},Y,\bar{Y}$; for convenience, we have explicitly extracted the leading $SU(N)$ colour scaling~$1/N^2$. In this way, we can compactly write the four-point functions of table~\ref{tab:fieldtheroycorr} below.

\subsection{Spin-chain interpretation}

\begin{figure}[t]
\begin{center}
\includegraphics[scale=0.3]{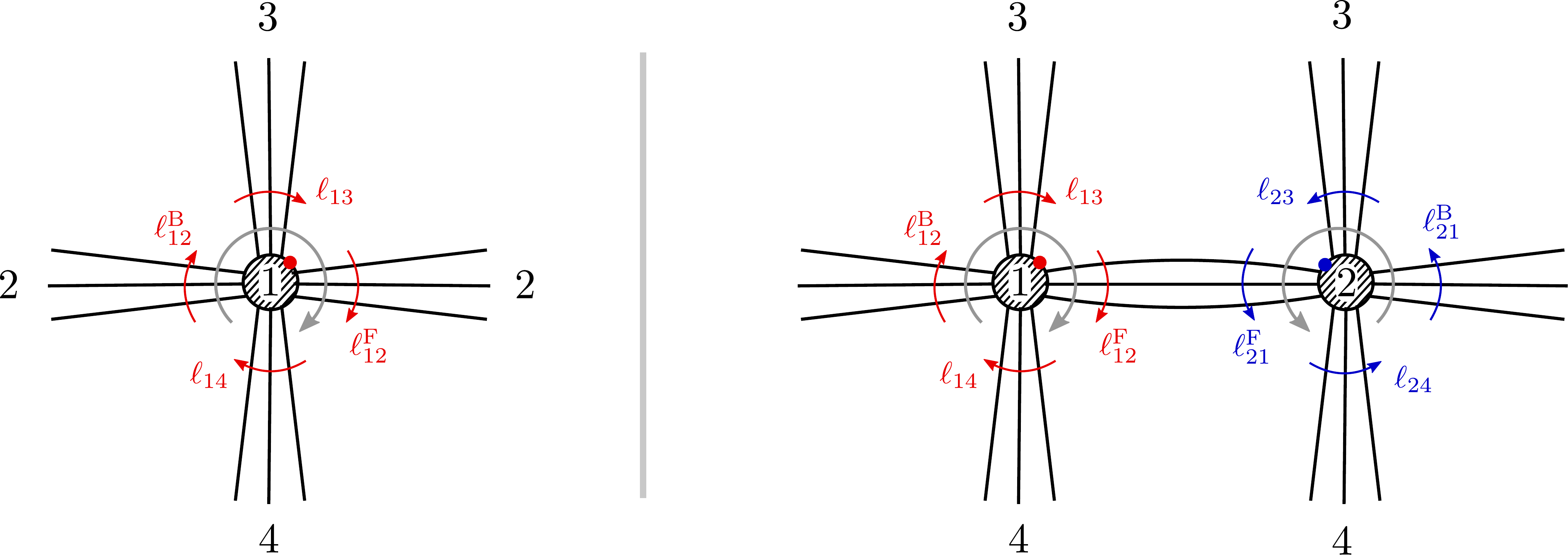}
\caption{Spin-chain picture for the spacetime dressing. We have one BMN operator on the left panel and two BMN operators on the right panel. We denote the number of propagators between two points $i,j$ as $\ell_{ij}$. When there are two possible ways for the Wick contraction, one in the front and one in the back, we denote the corresponding number of propagators by $\ell_{ij}^\text{F}$ and $\ell_{ij}^\text{B}$ respectively.
}
\label{fig:spinchain}
\end{center}
\end{figure}

The field-theory prescription for dressing the four-point function is rather straightforward. It is also interesting to obtain the same results in the spin-chain picture, which can be done explicitly at tree level. Let us start by presenting a neat argument originally given in ref.~\cite{Fleury:2016ykk}. We first consider the spacetime dressing for a single BMN operator and then extend the argument to the case with two BMN operators. Consider figure~\ref{fig:fourtessellations} where we place the BMN operator at position $a_1$. For simplicity, we take the one magnon state with impurity of flavour~$Y$ and momentum~$p$. At tree-level we find
\begin{equation}
\begin{aligned}
\label{eq:one_BMN}
\text{T}^{Y}_1=&
\sum_{n = 1}^{\ell_{12}^{\text{F}}} \frac{e^{i p n}}{a_{12}} + e^{i p\, \ell_{12}^{\text{F}}} \sum_{n=1}^{\ell_{14}} \frac{e^{i p n}}{a_{14}} + e^{i p (\ell_{12}^{\text{F}}+\ell_{14})} \sum_{n=1}^{\ell_{12}^{\text{B}}} \frac{e^{i p n}}{a_{12}}+e^{ip(\ell_{12}^{\text{F}}+\ell_{14}+\ell_{12}^{\text{B}})}\sum_{n=1}^{\ell_{13}}\frac{e^{ipn}}{a_{13}}\\
=&\frac{1}{\cN(p)} \left[v_{1;23}+v_{1;42}\,e^{ip\ell_{12}^{\text{F}}}+v_{1;24}\,e^{ip(\ell_{12}^{\text{F}}+\ell_{14})}
+v_{1;32}\,e^{ip(\ell_{12}^{\text{F}}+\ell_{14}+\ell_{12}^{\text{B}})}\right].
\end{aligned}
\end{equation}
where $v_{i;jk}=1/a_{ij}-1/a_{ik}$ and we have used $e^{ipL_1}=1$, ($L_1=\ell_{12}^{\text{F}}+\ell_{14}+\ell_{12}^{\text{B}}+\ell_{13}$).
This result reproduces exactly the ``position-dressed'' hexagon form factor $\widehat{\mathfrak{h}}_{i;jk}$ of eq.~\eqref{eq:hexagoncorr} up to the overall normalisation $\cN(p)=e^{-ip}-1$.

Things become more interesting when we consider \textit{two} BMN operators with non-trivial excitations. Let us consider an example where excitations on both operators are $Y$s. We get a term which is the product of two factors---one for each operator---which are similar to the one in eq.~(\ref{eq:one_BMN})
\begin{equation}
\begin{aligned}
\label{eq:TT_BMN}
\text{T}^{Y}_1=&\frac{1}{\cN(p)} \left[v_{1;23}+v_{1;42}\,e^{ip\ell_{12}^{\text{F}}}+v_{1;24}\,e^{ip(\ell_{12}^{\text{F}}+\ell_{14})}
+v_{1;32}\,e^{ip(\ell_{12}^{\text{F}}+\ell_{14}+\ell_{12}^{\text{B}})}\right],\\
\text{T}^{Y}_2=&\frac{1}{\cN(q)} \left[v_{2;13}+v_{2;41}\,e^{iq\ell_{21}^{\text{F}}}+v_{2;14}\,e^{iq(\ell_{21}^{\text{F}}+\ell_{24})}
+v_{2;31}\,e^{iq(\ell_{21}^{\text{F}}+\ell_{24}+\ell_{21}^{\text{B}})}\right].
\end{aligned}
\end{equation}
However, there is a contact term arising when we have $Y$-excitations at corresponding positions on both operators. The propagator vanishes in this case and we should subtract the corresponding contribution, which is
\begin{align}
\label{eq:CYY_contact}
\text{C}^{YY}_{12} =&\, \sum_{n=1}^{\ell_{12}^{\text{F}}} \frac{e^{ipn}}{a_{12}} \frac{e^{iqn}}{a_{21}}
+e^{ip(\ell_{12}^{\text{F}}+\ell_{14})}e^{ip(\ell_{21}^{\text{F}}+\ell_{24})}\sum_{n=1}^{\ell_{12}^{\text{B}}} \frac{e^{ipn}}{a_{12}} \frac{e^{iqn}}{a_{21}}\\\nonumber
=&\,-\frac{1}{a_{12}^2}\frac{1}{\mathcal{N}(p+q)}\left(1-e^{i(p+q)\ell_{12}^\text{F}}+
e^{ip(\ell_{12}^{\text{F}}+\ell_{14})+ip(\ell_{21}^{\text{F}}+\ell_{24})}(1-e^{i(p+q)\ell_{12}^\text{F}})  \right),
\end{align}
so that the corrected result is
\begin{equation}
\label{eq:PiYY}
\text{T}^{YY}_{12} = \text{T}^{Y}_1 \text{T}^{Y}_2 - \text{C}^{YY}_{12}\,.
\end{equation}
To extract the hexagon form factors for two excitations of type $Y$,$Y$ at tree level, we can look at for example the coefficient of $e^{i(p+q)\ell_{12}^{\text{F}}}$. Noticing that $v_{1;42}v_{2;41}=-1/a_{12}$, we have
\begin{align}
v_{1;42}\,v_{2;41}\,\left(1+\frac{\mathcal{N}(p)\mathcal{N}(q)}{\mathcal{N}(p+q)}\right)
=v_{1;42}\,v_{2;41}\,\frac{u-v-i}{u-v}\,,
\end{align}
where the second term in the bracket comes from the contact term and we have used the change of variables $e^{ip}=(u+i/2)/(u-i/2)$ and $e^{iq}=(v-i/2)/(v+i/2)$. Once again, up to normalisation, eq.~\eqref{eq:PiYY} reproduces the hexagon amplitude $\widehat{\mathfrak{h}}$ including the space-time dressing, \textit{cf.}\ eq.~\eqref{eq:hexagoncorr}. Here we considered two excitations of type $Y,Y$; the explicit form of (\ref{eq:TT_BMN}) and of the contact terms (\ref{eq:CYY_contact}) depend on the choice of excitations, \textit{cf.}\ eq.~\eqref{eq:propagators}. It is easy to check that the matching works more generally. A calculation similar to the one above gives $\text{T}_1^{X}=\text{T}_1^{\bar{X}}=0$ and $\text{T}_1^{\bar{Y}}=-\text{T}_1^{Y}$. The contact terms are given by
\begin{align}
\text{C}_{12}^{X\bar{X}}=\text{C}_{12}^{\bar{X} X}=-\text{C}_{12}^{\bar{Y}\bar{Y}}=-\text{C}_{12}^{YY}.
\end{align}
All the other contact terms are zero. The four-point functions are given by
\begin{align}
\text{T}^{\chi_1\chi_2}_{12}=\text{T}_1^{\chi_1}\text{T}_2^{\chi_2}-\text{C}_{12}^{\chi_1\chi_2},\qquad \chi_i=X,\bar{X},Y,\bar{Y}.
\end{align}
Moreover, since the excitations in each operator are magnons of an integrable spin chain, it is natural that this structure generalises to more complicated multi-excitation states.

\afterpage{%
    \clearpage
    \thispagestyle{empty}
    \begin{landscape}
\begin{table}[h]
\begin{center}
\vspace*{-1.75cm}
\begin{tabular}{|c|c|c|c|}
\hline
	Correlator & $\mathbf{m}_{Y,Y}$ & $\mathbf{m}_{Y,\bar{Y}}$ & $ \mathbf{m}_{X,\bar{X}}$ \\
	\hline
	$\braket{\mathcal{B}_4 \mathcal{B}_4 \mathcal{O}_2 \mathcal{O}_2}$ & $\frac{4}{3} (1,-2,2,-2,1)$ & $\frac{4}{3}(1,0,3,0,1)$ & $\frac{44}{3}(0,0,1,0,0)$  \\
	\hline
	$\braket{\mathcal{B}_4 \mathcal{B}_4 \mathcal{O}_4 \mathcal{O}_2}$ & $(3,-8,10,-8,3)$ & $(3,-4,6,-4,3)$ & $\frac{32\sqrt{2}}{3}(0,0,1,0,0)$  \\	
	$\braket{\mathcal{B}_4 \mathcal{B}_4 \mathcal{O}_3 \mathcal{O}_3}$ & $0$ & $4\sqrt{\frac{5}{3}}(0,0,1,0,0)$ & $16(0,0,1,0,0)$  \\
	$\braket{\mathcal{B}_5 \mathcal{B}_4 \mathcal{O}_3 \mathcal{O}_2}$ & $(3,-6,5,-4,2)$ & $(3,-2,4,0,2)$ & $19(0,0,1,0,0)$  \\
	$\braket{\mathcal{B}_5 \mathcal{B}_5 \mathcal{O}_2 \mathcal{O}_2}$ & $(3,-4,6,-4,3)$ & $(3,2,4,0,3)$ & $26(0,0,1,0,0)$  \\
	$\braket{\mathcal{B}_6^{\pm} \mathcal{B}_4 \mathcal{O}_2 \mathcal{O}_2}$ & $-\frac{3}{2}(\mp 7\sqrt{3}+\sqrt{15})(0,0,1,0,0)$ & $0$ & $-2\sqrt{3}(\pm 7+\sqrt{5})(0,0,1,0,0)$  \\
	\hline
	$\braket{\mathcal{B}_4 \mathcal{B}_4 \mathcal{O}_4 \mathcal{O}_4}$ & $\frac{4}{3}(4,-12,10,-12,4)$ & $\frac{16}{3}(1,-2,1,-2,1)$ & $\frac{40}{3}(0,0,1,0,0)$  \\	
	$\braket{\mathcal{B}_4 \mathcal{B}_4 \mathcal{O}_5 \mathcal{O}_3}$ & $0$ & $\frac{8\sqrt{2}}{3}(0,0,1,0,0)$ & $16\sqrt{\frac{5}{2}}(0,0,1,0,0)$  \\
	$\braket{\mathcal{B}_4 \mathcal{B}_4 \mathcal{O}_6 \mathcal{O}_2}$ & $0$ & $\frac{8}{\sqrt{3}}(0,0,1,0,0)$ & $\frac{8}{\sqrt{3}}(0,0,1,0,0)$  \\
	$\braket{\mathcal{B}_5 \mathcal{B}_4 \mathcal{O}_4 \mathcal{O}_3}$ & $(4,-10,11,-8,3)$ & $\sqrt{2}(4,-6,8,-4,3)$ & $17\sqrt{2}(0,0,1,0,0)$  \\
	$\braket{\mathcal{B}_5 \mathcal{B}_4 \mathcal{O}_5 \mathcal{O}_2}$ & $0$ & $5\sqrt{\frac{5}{3}}(0,0,1,0,0)$ & $17\sqrt{\frac{5}{3}}(0,0,1,0,0)$  \\
	$\braket{\mathcal{B}_5 \mathcal{B}_5 \mathcal{O}_3 \mathcal{O}_3}$ & $(9,-18,14,-18,9)$ & $(9,-6,15,-6,9)$ & $41(0,0,1,0,0)$  \\
	$\braket{\mathcal{B}_5 \mathcal{B}_5 \mathcal{O}_4 \mathcal{O}_2}$ & $\sqrt{2}(3,-6,10,-6,3)$ & $\sqrt{2}(3,-2,7,-2,3)$ & $29\sqrt{2}(0,0,1,0,0)$  \\
\hline
	$\braket{\mathcal{B}_6^\pm \mathcal{B}_4 \mathcal{O}_3 \mathcal{O}_3}$ &
	$\frac{3}{10}(-5\sqrt{3}+3\sqrt{15})\,(1,-2,\frac{-19-13\sqrt{5}}{3},-2,1)$ &  $\frac{1}{10}(-5\sqrt{3}+3\sqrt{15})\,(3,-2,5+\sqrt{5},-2,3)$ & $\frac{\sqrt{3}}{5}(-85+3\sqrt{5})(0,0,1,0,0)$  \\
	$\braket{\mathcal{B}_6^\pm \mathcal{B}_4 \mathcal{O}_4 \mathcal{O}_2}$ &
	$\frac{2}{15}(\mp 5\sqrt{6}+\sqrt{30})\,(2,-4,13\pm 3\sqrt{5},-2,1)$ & $\frac{1}{15}(\mp 5\sqrt{6}+\sqrt{30})\,(4,-4,9\pm \sqrt{5},0,2)$  & $-\frac{\sqrt{6}}{5}(65+3\sqrt{5})(0,0,1,0,0)$  \\
	$\braket{\mathcal{B}_6^{\pm} \mathcal{B}_5 \mathcal{O}_3 \mathcal{O}_2}$ &
	$\frac{1}{5}(\mp 5\sqrt{3}+\sqrt{15})\,(2,-3,\frac{35\pm 11 \sqrt{5}}{4},-2,3)$ &
	$\frac{1}{10}(5\sqrt{3}-\sqrt{15})\,(-4,2,\pm \sqrt{5},0,-3)$ & $\frac{\sqrt{3}}{5}(\pm90+11\sqrt{5})(0,0,1,0,0)$  \\
	$\braket{\mathcal{B}_6^+ \mathcal{B}_6^+ \mathcal{O}_2 \mathcal{O}_2}$ & \begin{tabular}{l}
	$\frac{4}{5}(3+\sqrt{5})\,(-1,1,\frac{-281-115\sqrt{5}}{8},1,-1)$\end{tabular} & \begin{tabular}{l}
	$\frac{4}{5}(3+\sqrt{5})\,(-1,0,-\frac{5}{3+\sqrt{5}},0,-1)$\end{tabular} & $\frac{\sqrt{2}}{5}(137+16\sqrt{5})(0,0,1,0,0)$  \\
	$\braket{\mathcal{B}_6^+ \mathcal{B}_6^- \mathcal{O}_2 \mathcal{O}_2}$ &
	$\frac{8}{5}(-1,1,\frac{43}{4},1,-1)$ &
	$\frac{8}{5}(-1,0,0,0,-1)$ & $\frac{86}{5}(0,0,1,0,0)$  \\
	$\braket{\mathcal{B}_6^- \mathcal{B}_6^- \mathcal{O}_2 \mathcal{O}_2}$ & \begin{tabular}{l}
	$\frac{4}{5}(3+\sqrt{5})\,(1,-1,\frac{281-115\sqrt{5}}{8},-1,1)$\end{tabular} & \begin{tabular}{l}
	$\frac{4}{5}(3+\sqrt{5})\,(1,0,-\frac{5}{3+\sqrt{5}},0,-1)$\end{tabular} & $\frac{\sqrt{2}}{5}(137+16\sqrt{5})(0,0,1,0,0)$  \\
\hline
	$\braket{\mathcal{B}_7' \mathcal{B}_4 \mathcal{O}_3 \mathcal{O}_2}$ & $6(0,0,1,0,0)$ & $0$ & $20(0,0,1,0,0)$  \\
	$\braket{\mathcal{B}_7'' \mathcal{B}_4 \mathcal{O}_3 \mathcal{O}_2}$ & $12\sqrt{3}(0,0,-1,0,0)$ & $0$ & $26\sqrt{3}(0,0,-1,0,0)$  \\
	$\braket{\mathcal{B}_7' \mathcal{B}_5 \mathcal{O}_2 \mathcal{O}_2}$ & $20(0,0,1,0,0)$ & $0$ & $40(0,0,1,0,0)$  \\
	$\braket{\mathcal{B}_7'' \mathcal{B}_5 \mathcal{O}_2 \mathcal{O}_2}$ & $12\sqrt{3}(0,0,-1,0,0)$ & $0$ & $24\sqrt{3}(0,0,-1,0,0)$  \\
\hline
	\end{tabular}
\caption{Tree-level connected four-point functions with two single-trace BMN operators with excitations of type $\chi$ on the first operator and $\chi'$ on the second, as computed by Wick contractions. Here $(\chi,\chi')=(Y,Y)$, $(Y,\bar{Y})$ or $(X,\bar{X})$, and we suppressed a factor of $1/N^2$, \textit{cf.}\ eq.~\eqref{eq:4pbasis}.}
\label{tab:fieldtheroycorr}
\end{center}
\end{table}
\end{landscape}
\clearpage
}

\subsection{Tessellating the four-point function}

\begin{figure}[t]
\begin{center}
\includegraphics[width=\linewidth]{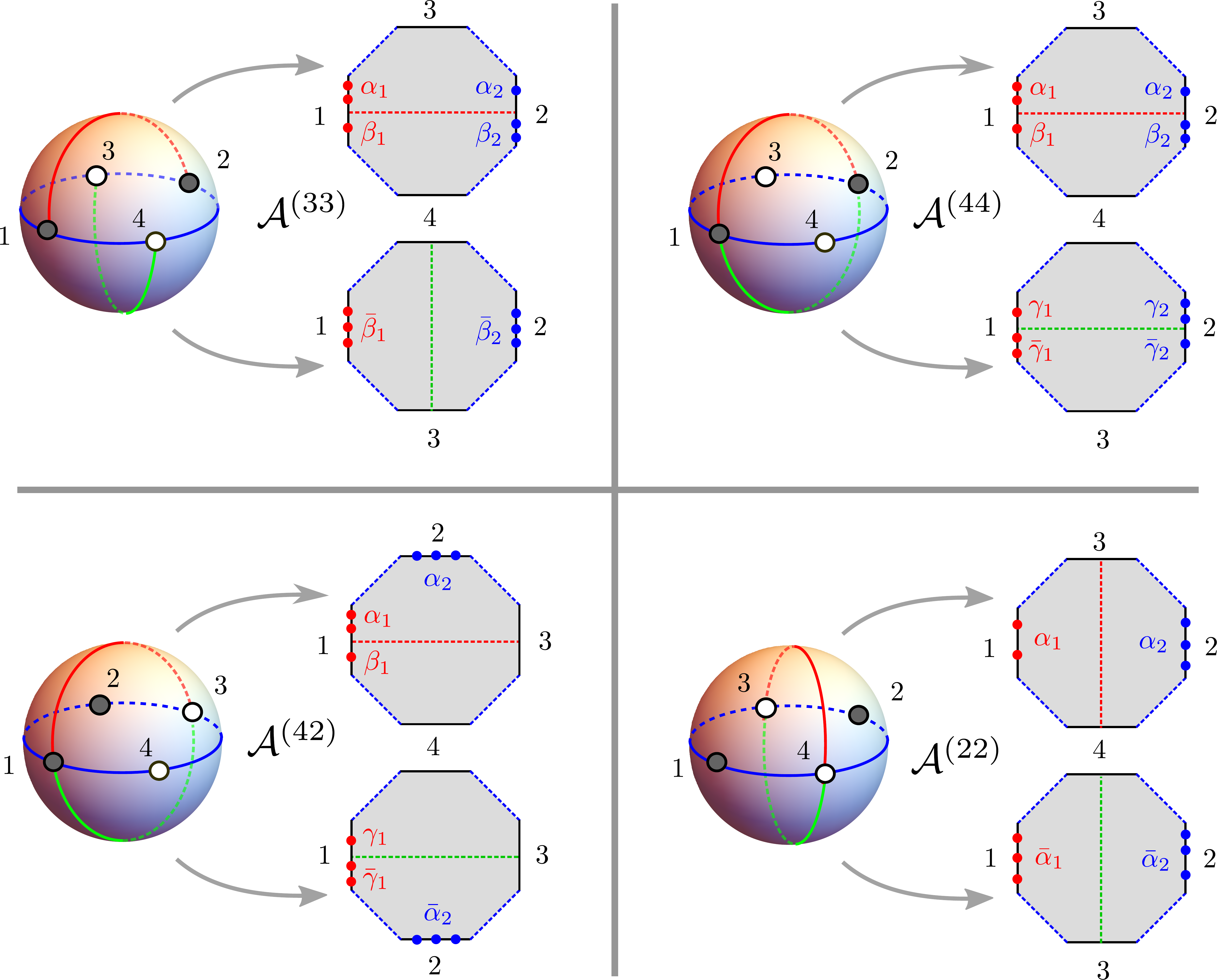}
\caption{Four tessellations of the sphere into hexagons.
The four-point function of two non-protected operators at positions $1,2$ (dark dots) and two $\tfrac{1}{2}$-BPS ones (white dots) is depicted on the sphere. Strands of $\ell_{ij}$ propagators are denoted as lines connecting points $i$ and~$j$. Planarly, only six set of ``edge-widths'' $\ell_{ij}$ can be non vanishing; this gives the four topologies of the picture.
For each of them, we find an hexagon tessellation by cutting the sphere along the edges with $\ell_{ij}>0$. We denote the corresponding hexagon amplitudes as $\mathcal{A}^{(kl)}$ to highlight that operator~$1$ has been cut into~$k$ pieces and operator~2 into $l$~pieces.
As always, the computation of $\mathcal{A}^{(kl)}$ requires summing over magnon partitions, which we denote by Greek letters.
}\label{fig:fourtessellations}
\end{center}
\end{figure}

We have seen how to incorporate the dependence on spacetime (and R-symmetry charges) in the hexagon; we still have to work out how to cut the four-point function into hexagons. There are two routes: obviously, we could first split the four-point functions into two three-point functions, like in the OPE, and tessellate those. This however requires summing over intermediate \textit{physical} states, including multi-trace operators, which is rather involved, see ref.~\cite{Basso:2017khq} for an implementation of this approach. What we advocated in ref.~\cite{Eden:2016xvg} is decomposing the hexagon along ``mirror cuts'', \textit{cf.}\ figure~\ref{fig:4pttwocuts}. A first question is exactly in which way the four-point function should be cut. In figure~\ref{fig:fourtessellations} we highlight four distinct ways to cut the four-point function of two BMN and two $\tfrac{1}{2}$-BPS operators along mirror lines. As explained in the caption, the position of the propagators naturally suggests how to tessellate the sphere.

There is an important subtlety, however. Diagrams that have vanishing bridge-lengths $\ell_{ij}=0$ for several choices of $i$ and $j$ may be represented over more than one of the topologies of figure~\ref{fig:fourtessellations}. This prompts the question of whether we should sum over these different topologies or we are free to pick the one that suits us best. In ref.~\cite{Eden:2016xvg} we proposed that the result of the tessellation approach is independent on how we cut the diagram, whenever we have multiple choices. We call this property \textit{embedding invariance}. This was checked explicitly over a number of tree-level examples~\cite{Eden:2016xvg}, and in ref.~\cite{Fleury:2016ykk} the property  was shown to hold also at one loop for a particular setup. In fact, in all the examples we will consider in this paper we find that embedding invariance holds, though we cannot yet offer a proof for it from general principles.

We now come to our recipe for computing the four-point functions. Firstly, we list all the diagrams which we expect from free field theory, by taking all possible planar Wick contractions. Next, we look at how these can be embedded in the hexagonal tessellations of figure~\ref{fig:fourtessellations}. If it is possible to choose more than one embedding, we are free to pick the one that suits us best.
One subtlety may arise in the embedding; consider topology (33): it may happen that the same diagram can be embedded in two inequivalent ways on that topology, if we have three non-vanishing edges, say, on operator~1. For instance, we could arrange them so that they connect to operators $3,2$ and~4, ordered clockwise; or we could arrange them so that they connect to $3, 4$ and~$2$ again clockwise, instead.  We call these graphs \textit{chiral}, and we count both such embeddings separately; clearly, a similar issue may arise for topology (44). Having listed the hexagon tessellations with appropriate multiplicities, the tree-level result can be found by evaluating the hexagon form factors, taking care of using the splitting factors $\omega(\alpha,\bar{\alpha},\ell_{ij})$ \eqref{eq:oldmathcalA} and the position-dependent hexagons~\eqref{eq:hexagoncorr}.

To illustrate the procedure, which was detailed in ref.~\cite{Eden:2016xvg}, let us spell out $\cA^{(22)}$, which is the least cumbersome diagram. We assume that the magnons of operator 1 are originally on the back of the figure, while those of operator 2 are on the front. We obtain
\begin{equation}
\begin{aligned}
\cA^{(22)} =
\sum_{\alpha\cup\bar{\alpha}=\{u_1,u_2\}} \sum_{\beta\cup\bar{\beta}=\{u_1',u_2'\}} \omega({\alpha},\bar{\alpha}, \ell_{13}) \, \omega({\beta},\bar{\beta}, \ell_{24})
\hspace*{3.5cm}\\
 \widehat{\mh}_{143}^{\text{B}}(\alpha,\emptyset,\emptyset) \, \widehat{\mh}_{134}^{\text{B}}(\bar{\alpha},\emptyset,\emptyset) \, \widehat{\mh}_{243}^{\text{F}}(\beta,\emptyset,\emptyset) \, \widehat{\mh}_{234}^{\text{F}}(\bar{\beta},\emptyset,\emptyset),
\end{aligned}
 \label{eq:ca22}
\end{equation}
where the labels B and~F distinguish the back and front of the figure for the convenience of the reader. To simplify our notation, we made the dependence of the hexagon operators on the space-time factors implicit, \textit{cf.}\ eq.~\eqref{eq:hexagoncorr}. The other cases in figure \ref{fig:fourtessellations} yield analogous expressions, but obviously with partitions into more sets. The full four-point function is given by summing over all diagrams, each counted once,
\begin{equation}
\label{eq:4ptformula}
\braket{\mathcal{B}_{L_1} \mathcal{B}_{L_2} \mathcal{O}_{L_3} \mathcal{O}_{L_4}}_c=\frac{1}{N^2}\sqrt{\frac{L_1L_2L_3L_4}{\mathcal{G}_1\mathcal{G}_2 S_{12}S_{34}}} \left[ \sum_{(jk) = (33),(44),(42), \ \underline{\ell}}  \mathcal{A}^{(jk)}_{\underline{\ell}} \right],
\end{equation}
where the subscript $c$ indicates that we are computing the \textit{connected} part of the correlator and $\mathcal{A}^{(jk)}$ are the hexagon amplitudes defined in ref.~\cite{Eden:2016xvg}. Notice that we removed topology~(22) from the sum; this is a minor simplification that is possible for the particular cases studied here---all diagrams can also be represented on the other three topologies, so that by embedding invariance we do not need to consider (22).

\subsection{Edge-reducible graphs}
\label{sec:edgered}
The rules that we have summarised above were successfully employed in ref.~\cite{Eden:2016xvg} to compute tree-level correlation functions involving one non-protected operator. As we discussed in section~\ref{sec:positiondressing} (see also appendix~\ref{app:position}), they coincide with the prescription of ref.~\cite{Fleury:2016ykk} for the asymptotic part of the hexagon correlators. An interesting observation was made in ref.~\cite{Fleury:2016ykk} when studying the one-magnon L\"uscher-like corrections to correlators (\textit{cf.}\ section~4.5 there): the authors propose that, in order to match field theory, one should sum over all connected graphs at the asymptotic level, and only over ``one-edge irreducible'' graphs for L\"uscher corrections---\textit{i.e.}\ over graphs that cannot be disconnected by cutting a single edge with however many propagators. We will see that this empirical rule fails at tree-level when considering four-point correlators involving two non-protected operators. As we will detail in the next section, we propose that the correct prescription is instead to dress graphs by their $SU(N)$ colour factor.

\section{Four-point functions with two non-protected operators by hexagons}
\label{sec:twoBMN}
The computation of tree-level four-point functions with any number of non-protected operators should follow straightforwardly from the general rules of the previous section. Nonetheless, it is worth detailing one such computation, as it will reveal an important subtlety in the hexagon formalism.

\subsection{One example and a puzzle: \texorpdfstring{$\langle \cB_5\cB_4\cO_3\cO_2 \rangle$}{<B5 B4 O3 O2>}}
\label{sec:puzzle}

In this section we work out in full detail one particular example among the correlation functions of table~\ref{tab:fieldtheroycorr}: $\langle \cB_5\cB_4\cO_3\cO_2 \rangle$.
To make our computation more explicit, we slightly alter eq.~\eqref{eq:4ptformula} by distinguishing the contributions of different graphs by coefficients $c^{jk}_{\underline{l}}$:
\begin{equation}
\braket{\mathcal{B}_{L_1} \mathcal{B}_{L_2} \mathcal{O}_{L_3} \mathcal{O}_{L_4}}_c=\frac{1}{N^2}\sqrt{\frac{L_1L_2L_3L_4}{\mathcal{G}_1\mathcal{G}_2S_{12}S_{34}}} \left[ \sum_{(jk) = (33),(44),(42), \ \underline{\ell}} c^{(jk)}_{\underline{\ell}}\, \mathcal{A}^{(jk)}_{\underline{\ell}} \right].
\end{equation}
\begin{table}[t]
\begin{center}
\begin{tabular}{l|l|l}
	Graph $\phantom{\Bigg|}$ & multiplicity $*\, \mathcal{A}^{(33)}_{\Pi_{12},\Pi_{13},\Pi_{24}}$ & multiplicity  $*\,\mathcal{A}^{(44)}_{\Pi^f_{12},\Pi^b_{12},\Pi_{13},\Pi_{23}}$
\\
\hline
	$(\Pi_{12})^2(\Pi_{13})(\Pi_{14})^2(\Pi_{23})^2$ & $2 c_1^{(33)}\  *\   \mathcal{A}^{(33)}_{2,1,0}$ & $\ c^{(44)}_{1}\ *\  \mathcal{A}^{(44)}_{1,1,1,2}$  \\
	$(\Pi_{12})^2(\Pi_{13})^2(\Pi_{14})(\Pi_{23})(\Pi_{24})$ & $ 2 c_2^{(33)}\  *\  \mathcal{A}^{(33)}_{2,2,1}$ & $\ c^{(44)}_{2}\ *\ \mathcal{A}^{(44)}_{1,1,2,1}$ \\
	$(\Pi_{12})^2(\Pi_{13})^3(\Pi_{24})^2$ & $\phantom{2} c_3^{(33)}\  *\ \mathcal{A}^{(33)}_{2,3,2}$ & $\ c^{(44)}_{3}\ *\ \mathcal{A}^{(44)}_{1,1,3,0}$ \\
	$(\Pi_{12})^3(\Pi_{13})(\Pi_{14})(\Pi_{23})(\Pi_{34})$ & $2 c_4^{(33)}\  *\  \mathcal{A}^{(33)}_{3,1,0}$   \\
	$(\Pi_{12})^3(\Pi_{13})^2(\Pi_{24})(\Pi_{34})$ & $\phantom{2} c_5^{(33)}\  *\  \mathcal{A}^{(33)}_{3,2,1}$  \\
	$(\Pi_{12})^4(\Pi_{13})(\Pi_{34})^2$ & $\phantom{2} c_6^{(33)}\  *\  \mathcal{A}^{(33)}_{4,1,0}$ \\
	\end{tabular}
\caption{Relevant graphs and the corresponding hexagon amplitudes for the correlator  $\langle \cB_5\cB_4\cO_3\cO_2 \rangle$. The graphs' topologies are grouped according to figure~\ref{fig:fourtessellations}. Notice that one choice of propagators~$\Pi_{ij}$ can be embedded in multiple topologies depending on how the progagators are distributed among the front and back of the sphere. For instance, in the first line, we can write $(\Pi_{12})^2(\Pi_{13})(\Pi_{14})^2(\Pi_{23})^2$ in two ways on topology (33) when the two $\Pi_{12}$-propagators are both on the front or both on the back of the sphere---hence the factor of two. When one is on the front and one is on the back, we have a single graph of topology (44).
}\label{tab:5432}
\end{center}
\vskip -0.25 cm
\end{table}
We have listed in table~\ref{tab:5432} which graphs can contribute to this four-point function. As it turns out, by using embedding invariance, we can restrict to topologies (33) and (44). The result that we expect from free field theory can be found in table~\ref{tab:fieldtheroycorr}, and reads for impurities of type~$Y$,
\begin{equation}
\begin{aligned}
&\braket{ \mathcal{B}_5 \mathcal{B}_4 \mathcal{O}_3 \mathcal{O}_2}_c=\\
&\qquad3 \, v_{1;24}^2 v_{2;13}^2-6 \, v_{1;23} v_{1;24} v_{2;13}^2+5 \, v_{1;23}^2 v_{2;13}^2-4 \, v_{1;23}^2 v_{2;13} v_{2;14}+2 \, v_{1,23}^2 v_{2;14}^2\,.
\end{aligned}
\end{equation}
From the hexagon tessellation, keeping explicit the coefficients $c^{(33)}_{j}$ with $j=1,\dots 6$ and $c^{(44)}_{k}$ with $k=1,\dots 3$, we find:
\begin{equation}
\begin{alignedat}{2}
\braket{ \mathcal{B}_5 \mathcal{B}_4 \mathcal{O}_3 \mathcal{O}_2}= &(2\,c^{(33)}_{1}+c^{(44)}_{1}) \, v_{1;24}^2 v_{2;13}^2 \\
 +
&(-4 \, c^{(33)}_{1}+6 \, c^{(33)}_{2}-4 \, c^{(33)}_{4}-2 \, c^{(44)}_{1}-2 \, c^{(44)}_{2}) \, v_{1;23}v_{1;24}v_{2;13}^2 \\
 +
&\Big(2\,c^{(33)}_{1}-12 \, c^{(33)}_{2}+c^{(33)}_{3}+4 \, c^{(33)}_{4}+ 4 \, c^{(33)}_{5}\\
&\qquad\qquad\qquad+6 \, c^{(33)}_{6}+c^{(44)}_{1}+4 \, c^{(44)}_{2}+c^{(44)}_{3} \Big) \, v_{1;23}^2 v_{2;13}^2 \\
 +&
(6 \, c^{(33)}_{2}-2 \, c^{(33)}_{3}-4 \, c^{(33)}_{5}-2 \, c^{(44)}_{2}-2 \, c^{(44)}_{3}) \,v_{1;23}^2v_{2;13}v_{2;14} \\
 +
&(c^{(33)}_{3}+c^{(44)}_{3}) \, v_{1;23}^2v_{2;14}^2
\end{alignedat}
\end{equation}
Equating the two results and solving the resulting system we find:
\begin{equation}
\begin{alignedat}{2}
c^{(44)}_{1}&=3-2\,c^{(33)}_{1} \, , \\
c^{(44)}_{2}&=3 \, c^{(33)}_{2} - 2\,c^{(33)}_{4} \, , \\
c^{(44)}_{3}&=2-c^{(33)}_{3} \, ,\\
c^{(33)}_{5}&= c^{(33)}_{4} \, , \\
c^{(33)}_{6}&=0 \, .
\end{alignedat}
\end{equation}
Na\"ively, we would be tempted to set all coefficients $c^{(kk)}_{l}=1$, \textit{i.e.}\ to count once all the distinct graphs. This \textit{almost} works, \textit{except} from the condition $c^{(33)}_{6}=0$. Even if this specific test of the hexagon approach does not fix all coefficients, we find rather explicitly that one particular connected graph should be excluded. This graph is the only one-particle reducible  graph encountered in this case---though it is not the only one-\textit{egde} reducible example, as the graph corresponding to $c^{(33)}_{3}$ can be disconnected by cutting two propagators along a single edge. In the next section we will propose a systematic way to make sense of this discrepancy.

\begin{figure}[t]
\begin{center}
\includegraphics[scale=0.3]{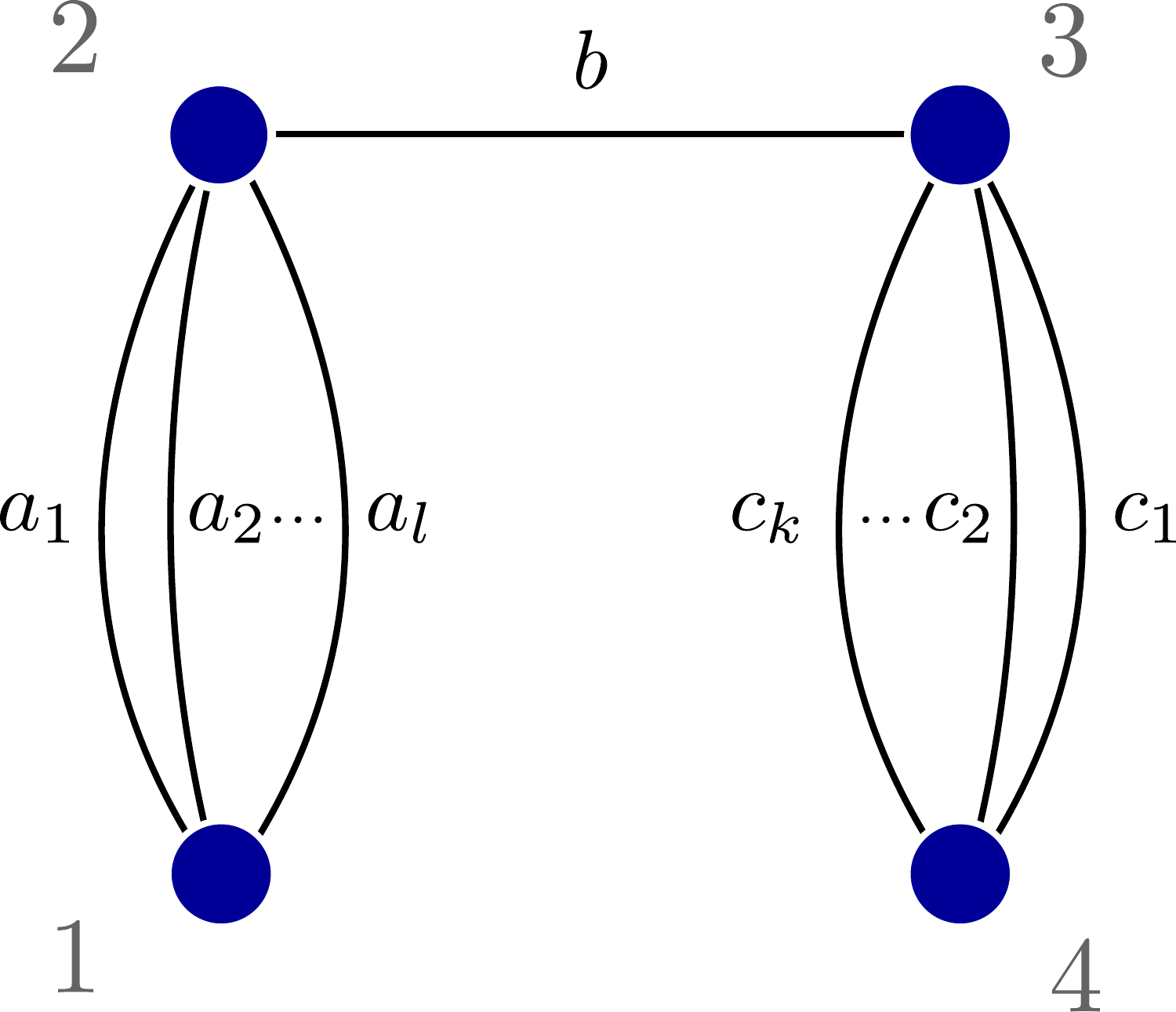}
\caption{A one-particle reducible four-point function. The lengths of the four operators that yield this graph are given by $L_1$, $L_2=L_1+1$, $L_3=L_4+1$ and $L_4$.
}\label{fig:1preducible}
\end{center}
\end{figure}

\subsection{Colour-dressing for hexagons}
\label{sec:colourdressing}
We propose that when computing the  diagrams of table~\ref{tab:5432} by the hexagon approach \emph{we should weight every diagram by its $SU(N)$ colour pre-factor}.
This rule allows us to reproduce all of the results of table~\ref{tab:fieldtheroycorr}, and modifies the original prescription in a non-trivial way, as highlighted by the example in the previous subsection. To further demonstrate this point, consider a one-particle reducible diagram such as the one of figure~\ref{fig:1preducible}. In general, this diagram gives a non-vanishing hexagon amplitude. However, its contribution vanishes in field theory. To see it, let us work out the colour part of the  Feynman diagram. Indicating the colour generator for line $a_1$, \textit{etc.}\ as $T_{a_1}$, \textit{etc.}, we get schematically
\begin{equation}
\begin{aligned}
&\text{Tr}\big[T_{a_1}\cdots T_{a_{l}}\big]\,
\text{Tr}\big[T_{a_{l}}\cdots T_{a_1}T_{b}\big]\,
\text{Tr}\big[T_{b}T_{c_1}\cdots T_{c_{k}}\big]\,
\text{Tr}\big[ T_{c_{k}}\cdots T_{c_1}\big]\\
&\hspace*{6cm}
\propto
\text{Tr}\big[T_{b}\big]\,
\text{Tr}\big[T_{b}T_{c_1}\cdots T_{c_{k}}\big]\,
\text{Tr}\big[ T_{c_{k}}\cdots T_{c_1}\big],
\end{aligned}
\end{equation}
where we used the well-known identities for $SU(N)$ generators, \textit{cf.}\ appendix~\ref{app:colour}, on the indices $a_1,\dots a_l$. Hence, the result vanishes as it is proportional to the trace of a single $SU(N)$ generator. For all the correlators involving a term like in figure~\ref{fig:1preducible} we find that \textit{such a diagram must be set to zero} in order to reproduce the correct correlation function by the hexagon approach. The reason why this subtlety might have been missed in earlier studies is that graphs of the type of figure~\ref{fig:1preducible} have a vanishing hexagon amplitude when considering a single non-protected operator.

Let us remark that, unlike the prescription for ``edge-reducible'' graphs of ref.~\cite{Fleury:2016ykk}, the constraint that we have found here applies already \textit{at tree-level}. This does not mean that colouring the hexagon formalism will not affect higher orders too. In fact, as we will discuss in section~\ref{sec:Nis2}, colouring plays a crucial role also at one loop, and reproduces the ``no edge-reducible'' empirical rule. We will also test our prescription for a rather involved setup, where we consider the leading $1/N$ corrections to a class of two-point functions---again, colouring is instrumental in recovering the field-theory result.
Finally, this whole line of reasoning suggests that there is a rather direct map between each diagram appearing in the hexagon tessellation and the graphs of field theory; it is interesting to explore how precise such a link may be; we will turn to this issue too in section~\ref{sec:Nis2}.

\subsection{Sub-extremal correlators and multi-trace admixtures}
\label{sec:subextremal}

\begin{table}[t]
\begin{center}
\begin{tabular}{l|l|l}
	Graph $\phantom{\Bigg|}$ &multiplicity$\,*\mathcal{A}^{(33)}_{\Pi_{12} \,\Pi_{13}\,\Pi_{24}}$ &multiplicity $*\mathcal{A}^{(44)}_{\Pi^F_{12}\,\Pi^B_{12}\,\Pi_{13}\,\Pi_{23}}$ \\
	\hline
	$(\Pi_{12})^4(\Pi_{13})(\Pi_{14})^2(\Pi_{23})$ &
	 $2\,c_1^{(33)}\!*\mathcal{A}^{(33)}_{410}$ &
$c^{(44)}_{11}*\mathcal{A}^{(44)}_{1311};\ c^{(44)}_{12}\!* \mathcal{A}^{(44)}_{2211};\ c^{(44)}_{13}\!*\mathcal{A}^{(44)}_{3111}$  \\
	$(\Pi_{12})^4(\Pi_{13})^2(\Pi_{14})(\Pi_{24})$ &
	 $2\,c_2^{(33)}\!*\mathcal{A}^{(33)}_{510}$ &  \\
	$(\Pi_{12})^5(\Pi_{13})(\Pi_{14})(\Pi_{34})$ &
	$2\,c_3^{(33)}\!*\mathcal{A}^{(33)}_{421}$ &
$c^{(44)}_{31}\!*\mathcal{A}^{(44)}_{1320};\ c^{(44)}_{32}\!*\mathcal{A}^{(44)}_{2220};\ c^{(44)}_{33}\!*\mathcal{A}^{(44)}_{3120}$  \\
	\end{tabular}
\caption{Relevant graphs and the corresponding hexagon contributions for the sub-extremal four-point function $\braket{ \mathcal{B}_7' \mathcal{B}_5 \mathcal{O}_2 \mathcal{O}_2}$.}
\label{tab:subextremal}
\end{center}
\end{table}

Sub-extremal four-point functions yield another example of correlators with interesting properties. We call a four-point function sub-extremal when the lengths of the four operators obey $L_1+2=L_2+L_3+L_4$. It is easy to work out that they contain three different types of graphs. In table~\ref{tab:fieldtheroycorr}, the sub-extremal cases are  $\braket{ \mathcal{B}_6^{\pm} \mathcal{B}_4 \mathcal{O}_2 \mathcal{O}_2}$,  $\braket{ \mathcal{B}_7 \mathcal{B}_4 \mathcal{O}_3 \mathcal{O}_2}$ and $\braket{ \mathcal{B}_7 \mathcal{B}_5 \mathcal{O}_2 \mathcal{O}_2}$, where in the last two correlators $\mathcal{B}_7'$ and $\mathcal{B}_7''$ can be used. Let us work out in detail the case of $\braket{ \mathcal{B}_7' \mathcal{B}_5 \mathcal{O}_2 \mathcal{O}_2}$. The possible graphs and their hexagon amplitudes are listed in table~\ref{tab:subextremal}.
The field theory result from table~\ref{tab:fieldtheroycorr} reads simply
\begin{equation}
\braket{ \mathcal{B}_7 \mathcal{B}_4 \mathcal{O}_2 \mathcal{O}_2}_c=20 \, v_{1;23}^2 v_{2;13}^2  \, ,
\label{eq:7522ft}
\end{equation}
while the hexagon tessellation yields
\begin{equation}
\begin{alignedat}{2}
\braket{ \mathcal{B}_7' \mathcal{B}_5 \mathcal{O}_2 \mathcal{O}_2}_c=
&(-6 \, c^{(33)}_{1}+c^{(44)}_{11}+4 \, c^{(44)}_{12}+c^{(44)}_{13}) \, v_{1;23}v_{1;24}v_{2;13}^2 \\
+ &(6 \, c^{(33)}_{1}-c^{(44)}_{11}+c^{(44)}_{12}-c^{(44)}_{13}\\
&\qquad\qquad\qquad+ 10 \, c^{(33)}_{2}+6 \, c^{(33)}_{3}-c^{(44)}_{31}+c^{(44)}_{32}-c^{(44)}_{33}) \, v_{1;23}^2 v_{2;13}^2 \\+
&(-6 \, c^{(33)}_{3}+c^{(44)}_{31}+4 \, c^{(44)}_{32}+c^{(44)}_{33}) \, v_{1;23}^2v_{2;13}v_{2;14} \,.
\end{alignedat}
\label{eq:7522hex}
\end{equation}
By comparing the two expression we see that setting all coefficients to one  gives a perfect matching.

In a sense, this perfect matching is bemusing, because for such a correlator we might expect leading-order contributions by double-trace admixtures. This was not the case for the example of section~\ref{sec:puzzle}, which was $\langle \cB_5\cB_4\cO_2\cO_2\rangle$; in fact by group theory a non-protected operator (``long'', from the point of view of $psu(2,2|4)$ representations) cannot mix with multi-trace operators involving only $\tfrac{1}{2}$-BPS single-trace components (which sit in ``short'' multiplets). This rules out any mixing for $\cB_4$ and $\cB_5$. However, longer operators \emph{can mix with double-trace operators}. An explicit diagonalisation of the one-loop dilatation operator confirms this. The eigenvalue problem for length 6 and 7 leads to complicated root functions of $N$, which we can expand at $N\gg1$. For instance, for
 $\cB_7'$ we have
\begin{equation}
\label{eq:B7primeadmixt}
\cB_7' - \frac{1}{N} \left( \frac{3}{2 \sqrt{2}} \, \cO_3 \, \cB_4 + 2 \sqrt{2} \, \cO_2 \, \cB_5 \right) + \dots
\quad \text{with}\quad
\gamma_1=2 + \frac{1}{N^2} \, \frac{11}{2} + \ldots ,
\end{equation}
for which we only indicated the leading and next-to leading orders in the $1/N$ expansion.
Notice that, as expected from $psu(2,2|4)$ representation theory, and given the small length of the operator, the mixing that we find has the form $\cO\cdot \cB$.

These admixtures potentially change the analysis of the sub-extremal  correlators which we consider, as they may contribute at the same order as the connected four-point functions of the single-trace parts. For instance, for the case of $\la \cB_7(a_1)' \cB_5(a_2) \cO_2(a_3) \cO_2(a_4) \ra_c$ which we considered above, we have a contribution due to admixtures at leading order, \textit{i.e.}\ at $O(1/N^2)$. Namely, we find
\begin{equation}
\begin{aligned}
&-\frac{2 \sqrt{2}}{N} \Big(\frac{1}{2} \la \cB_5(a_1) \cB_5(a_2) \ra \la \cO_2(a_1) \cO_2(a_3) \cO_2(a_4) \ra \\
&\qquad\qquad\quad+  \la \cO_2(a_1) \cO_2(a_3) \ra \la \cB_5(a_1) \cB_5(a_2) \cO_2(a_4) \ra + (3 \leftrightarrow 4) \Big) = -\frac{16 \, \kappa_{\chi,\chi'}}{N^2 \, a_{12}^4}\,,
\end{aligned}
\end{equation}
where the coefficient $\kappa_{\chi,\chi'}$ distinguishes the case where the excitations are of type $Y$ or $\bar{Y}$ or $X,\bar{X}$ on the first and second operators: $\kappa_{Y,Y}=1$, $\kappa_{\bar{Y},\bar{Y}}=0$ and $\kappa_{X,\bar{X}}=2$. Similar results hold \textit{e.g.}\ for $\la \cB_7'(a_1) \cB_4(a_2) \cO_3(a_3) \cO_2(a_4) \ra$ and   $\la \cB_6^\mp(a_1) \cB_4(a_2) \cO_2(a_3) \cO_2(a_4) \ra$.

In all the cases we consider, we find that the hexagon amplitudes are tailored to the single-trace part of the full conformal field theory eigenstates.
 This is not surprising, because the integrable spin-chain of $\mathcal{N}=4$ SYM is inherently single-trace. Indeed our diagrams on the sphere only account for one trace---\textit{i.e.}\ one puncture---at each point.
In the next section we will further explore how to use the hexagon formalism to compute two-point functions involving single-trace and double-trace operators; by a similar approach, we will reproduce two-point functions when the worldsheet has the topology of a torus.

\section{Hexagons and \texorpdfstring{$1/N$}{1/N} corrections}
\label{sec:1overN}

In the above analysis of four-point functions involving two non-protected operators we have learned two important lessons: firstly, we need to \emph{colour-dress the hexagon}  to know which diagrams we should take into account (\textit{cf.}\ section~\ref{sec:colourdressing}); secondly, that the hexagon amplitudes $\mathcal{A}^{(jk)}$  describe \emph{the single-trace part} of the conformal eigenstates, while double-trace and higher admixtures would eventually have to be dealt with separately (\textit{cf.}\ section~\ref{sec:subextremal}). In this section we will see that, keeping colour-dressing in mind, we can indeed reproduce the first correction to the norm of a conformal eigenstate. This comes with a relative factor of $1/N^2$ with respect to the leading order.

We shall focus on two-point functions giving the norm of a BMN operator~$\cB$ with two impurities including their admixtures. We will consider two cases: the two-point function between an operator $\cB$ with two impurities $X,X$ and its conjugate with two impurities $\bar{X},\bar{X}$, and the case where all impurities on both operators have flavour~$Y$. This last case might appear confusing from field-theory point of view; however, as detailed in ref.~\cite{Basso:2015zoa} this is the correct flavour identification in the hexagon formalism when the crossing transformation is accounted for. At any rate, these two computations should match for any given operator, due to $SU(4)$ symmetry. This is however not explicit in the hexagon formalism, and therefore performing both calculations will be a further check of our approach.
We hence consider the tree-level two-point functions of two conformal eigenstates with $SU(4)$ charges as above; these have a single-trace part and multi-trace admixtures.
We are interested in the two-point function up to order $1/N^2$, so that only the single-trace and the double-trace part are relevant:
\begin{equation}
\label{eq:twoptnonplanar}
\begin{aligned}
& \left\la \Big(\cB_n + \frac{c}{N} \cB_{n-m} \cO_m + \ldots\Big) \Big(\cB_n + \frac{c}{N} \cB_{n-m} \cO_m + \ldots\Big) \right\ra =\\
&\qquad\qquad \la \cB_n \cB_n \ra + \frac{2 c}{N} \la \cB_n (\cB_{n-m} \cO_m) \ra + \frac{c^2}{N^2} \la \cB_{n-m} \cB_{n-m} \ra \la \cO_{m} \cO_{m} \ra + O(N^{-4}).
\end{aligned}
\end{equation}
Remark that the mixing coefficient $c$ needs to be determined by diagonalising the finite-$N$ dilatation operator. This is a very non-trivial task, which falls outside the scope of this paper but that would be interesting to study by integrability---at least in a perturbativve $1/N$ expansion. Let us now look at eq.~\eqref{eq:twoptnonplanar} more closely.
The \emph{single-trace--single-trace} term is leading by construction. This terms contains several contributions that can be expanded in powers of $1/N^2$. The leading contribution comes from Wick contractions that can be represented on a sphere. How this can be computed by hexagons was described in appendix~K of ref.~\cite{Basso:2015zoa}; there, the authors recover the the off-shell scalar product (whence the Gaudin norm follows) from tessellating a three-point function  where the excitations on two operators are ``transverse'' and the third operator is $\tfrac{1}{2}$-BPS. Diagrams that can be drawn on a torus appear at order $1/N^2$, and we compute them in section~\ref{sec:torusnorm} by considering a \emph{four-point function} with two identity insertions.

Next, the \emph{single-trace--double-trace} contribution comes with an explicit $1/N$ pre-factor, and is further suppressed by $1/N$ due to the colour structure, so that all in all it comes at order $1/N^2$.
We will explain in section~\ref{sec:singletrdoubletr} how to compute the leading order of this correlation function from our four-point hexagon amplitude $\cA^{(42)}$ by point-splitting the double-trace operator and inserting one identity operator.

Finally, the \emph{double-trace--double-trace} term in \eqref{eq:twoptnonplanar} requires no further discussion: the disconnected term will be leading, and at leading order it will trivially give $c^2/N^2$ if the admixtures are written in terms of appropriately normalised operators.

\subsection{Two-point function on a torus}
\label{sec:torusnorm}

In figure~\ref{fig:torus} we have depicted the two-point function of two single-trace operators on a torus. With respect to a sphere, now we can draw ``planarly'', \textit{i.e.}\ without self-intersections, several strands of propagators which travel across the square's edges. In what follows, we will be specifically interested in those diagrams that can be drawn on the torus, but not on a sphere.

\subsubsection*{Colour factors}

\begin{figure}[t]
\includegraphics[width=\linewidth]{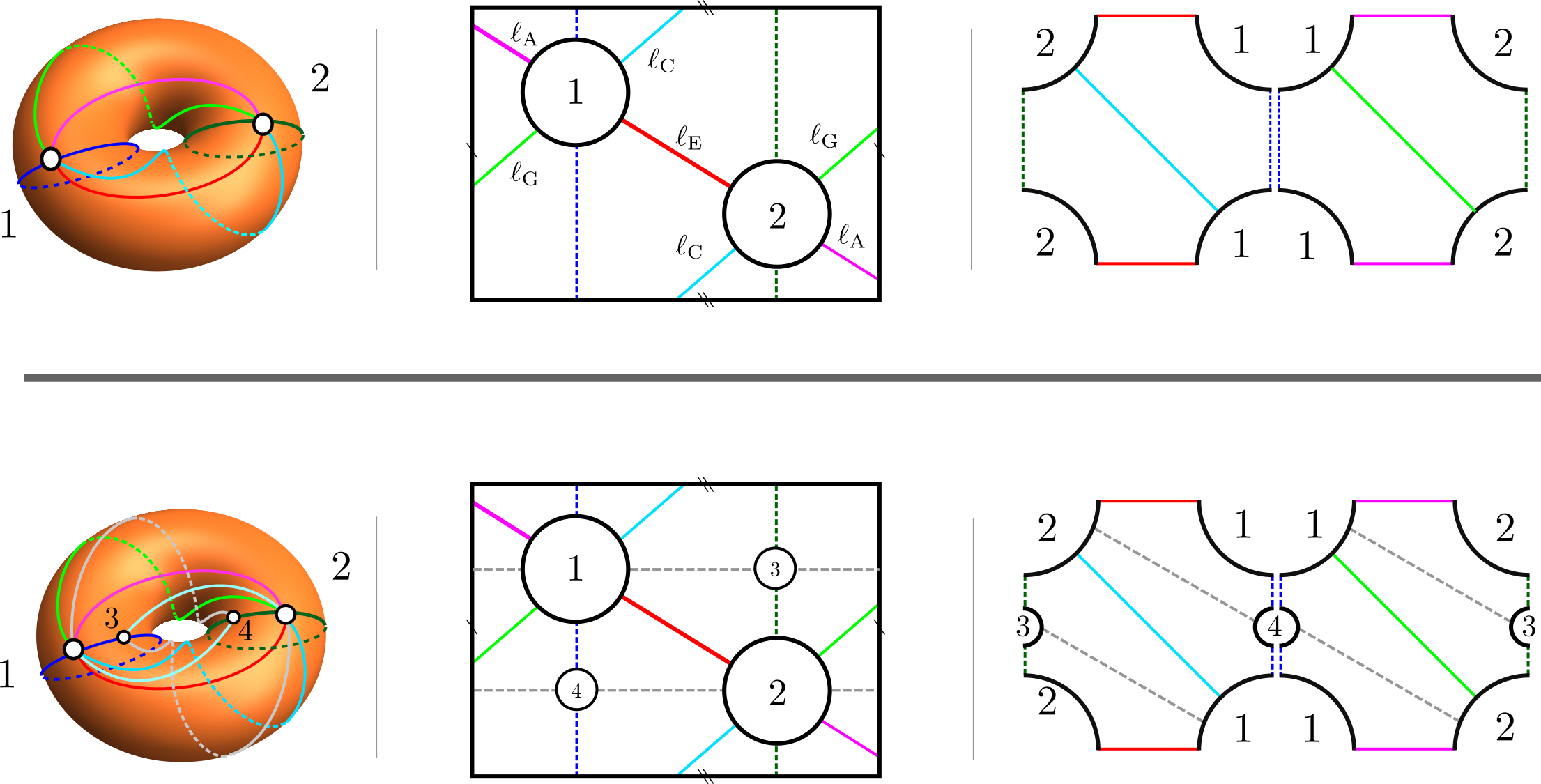}
\caption{Torus two-point function and hexagon tessellations. On the top left diagram, we draw a two-point function on the torus, highlighting all the possible propagators that can be drawn without self-intersections. Next, we represent this on a square with opposite edges identified. To cut it into four hexagons (rightmost panel), we follow the propagators, and include one additional cut which goes from each operator to itself, wrapping the A-cycle on the torus.
Below, we consider the same picture, but now we also introduce two operators labelled~$3,4$, obtaining the topology studied in ref.~\cite{Beisert:2002bb}. Here the additional operators  are needed to regularise the hexagon amplitude, and are taken to be the identity.
} \label{fig:torus}
\end{figure}

Our first goal is to determine the colour-factors for the torus diagrams. From the top-middle panel of figure~\ref{fig:torus} it is easy to do so. Let us index the four sets of propagators as $\{a_1,\ldots,a_{\ell_A}\}$, $\{c_1,\ldots,c_{\ell_C}\}$, $\{e_1,\ldots,e_{\ell_E}\}$ and $\{g_1,\ldots,g_{\ell_G}\}$; the $SU(N)$ generator corresponding to a field connected to the propagator $a_j$ will be denoted as $T_{a_j}$, and so on. By going around the operators $1$ and~$2$ and minding the ordering of each set of propagators, we read off the colour factor
\begin{equation}
\label{eq:Tcolour}
\begin{aligned}
&T_{\ell_A \ell_C \ell_E \ell_G} =
\text{Tr}\Big[
T_{a_1} \cdots T_{a_{\ell_A}} \, T_{c_1} \cdots T_{c_{\ell_C}}\, T_{e_1} \cdots T_{e_{\ell_E}}\, T_{g_1} \cdots T_{g_{\ell_G}}\Big] \\
& \qquad\qquad\qquad\qquad\text{Tr}\Big[
T_{a_{\ell_A}} \cdots T_{a_1} \, T_{c_{\ell_C}} \cdots T_{c_1} \, T_{e_{\ell_E}} \cdots T_{e_1} \, T_{g_{\ell_G}} \cdots T_{g_1}\Big].
\end{aligned}
\end{equation}
Imagining that each edge is a ribbon consisting of several propagators, it is easy to understand how the colour generators should be ordered in each trace by looking at how the ribbons are attached to the two operators. In particular, the generators of each ribbon should be sorted in opposite order in the two traces, \textit{cf.}\ figure~\ref{fig:torus}.  Due to the cyclicity of the trace it follows immediately that a colour factor $T_{\ell_A \ell_C \ell_E \ell_G}$ where two or more edge-widths $\ell_A, \ell_C, \ell_E,$ or  $\ell_G$ vanish can be mapped to the one on a sphere, \textit{i.e.}\ to $T_{\ell 000}$, where $\ell$ is the sum of the non-vanishing edge-widths. We have seen that colour factors are important for reproducing four-point functions by hexagons; we will see that this is the case also for two-point functions on the torus. In table~\ref{tab:torus} we have collected the evaluation of the diagrams needed for computing the torus two-point function by Wick contractions. We will discus the detail of that table in the next subsection; it is worth noting that certain classes of colour factors, namely $T_{ij10}$ for $i,j>1$ do not contribute at leading order (see appendix~\ref{app:colour} for the relevant $SU(N)$ manipulations) while others contribute with a~sign.

\subsubsection*{Tessellating the torus}

The simplest tessellation of the torus two-point function is given in the top-right panel of figure~\ref{fig:torus}; the torus is split over four hexagons. If we try to incorporate the space-time dependence of the hexagon, however, we immediately encounter an issue. Let us focus on the leftmost hexagon in the figure. Denoting such a hexagon as $h_{212}$, and inserting a single $Y$ excitation, we would find by our prescription~\eqref{eq:hexagoncorr}
\begin{equation}
\widehat{\mathfrak{h}}_{212}(\{u\},\emptyset,\emptyset)=
v_{2;12}\,\mathfrak{h}_{212}(\{u\},\emptyset,\emptyset)=
\Big(\frac{1}{a_{21}}-\frac{1}{0}\Big)\
\mathfrak{h}_{212}(\{u\},\emptyset,\emptyset),
\end{equation}
which clearly makes no sense, as the hexagon amplitude itself is non-vanishing and could not cancel such a pole.%
\footnote{%
The situation improves a bit if we consider the transverse excitations $X,\bar{X}$: then, the single-magnon hexagon form factor vanishes, and for two excitations we find a regular position-dependence owing to the identity $v_{1;12}v_{2;11}=1/a_{12}^2$. Still, we take the presence of divergences for longitudinal magnons as a sign that the set-up need to be regularised.
Besides, while it may be possible to do without any regularisation at three level for some flavours, it might be impossible to avoid this when considering wrapping effects.
}
 To remedy this pathology, we introduce two \textit{identity operators} at positions $a_3,a_4$ along the mirror edges that would lead to a self-contraction. Notice that these mirror edges have vanishing bridge-length. The resulting tessellation involves \textit{eight hexagons} and is presented in the lower panel of figure~\ref{fig:torus}. This picture is reminiscent of what was found in ref.~\cite{Beisert:2002bb} in the study of torus correlators.%
\footnote{We are grateful to Niklas Beisert for bringing ref.~\cite{Beisert:2002bb} to our attention%
}%
. This allows us to obtain finite results, but introduces a spurious dependence  on $a_3,a_4$; it will be a test of our construction that this dependence should drop out, and that the correlator should scale as $1/a_{12}^4$ as expected from free field theory.

\begin{figure}[t]
\begin{center}
\includegraphics[width=\linewidth]{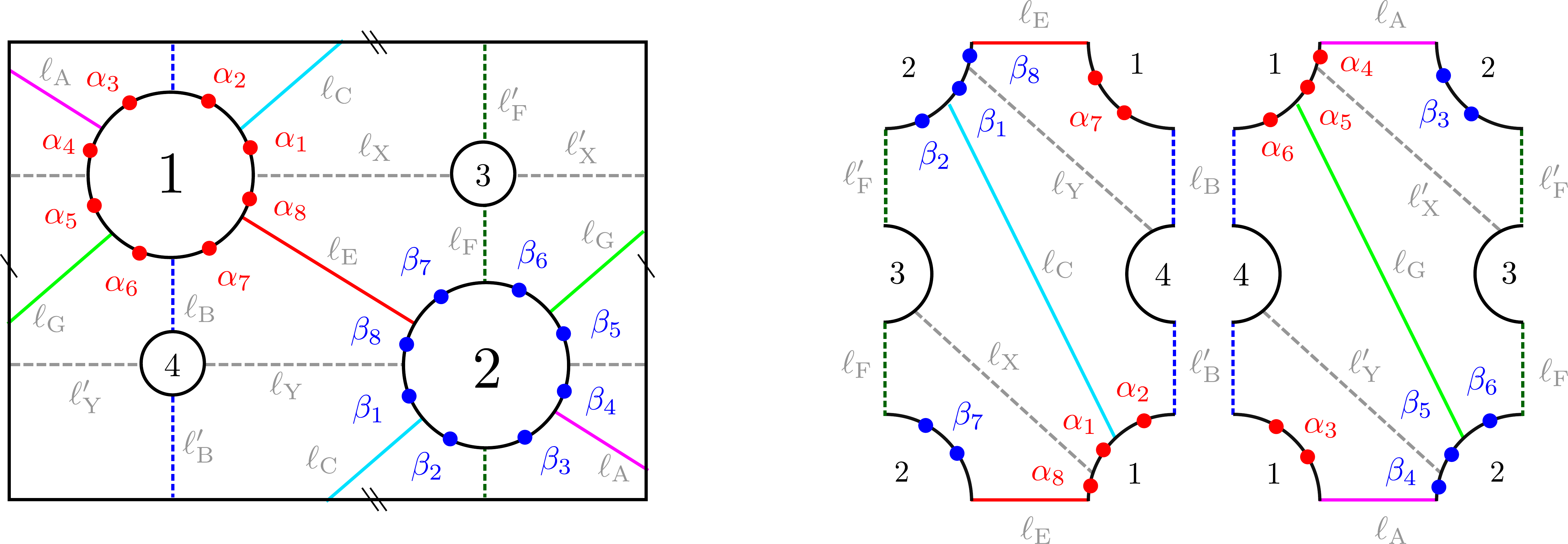}
\caption{We further detail the tessellation of the torus into eight hexagons. On the left, we draw the torus as a square; we insert two non-protected operators~$1,2$, and tessellate the torus into four hexagons by cutting along the propagators (solid coloured lines) and along one of the torus' cycles (coloured dotted lines). To regularise the construction we insert two identity operators labelled by $3,4$ along the coloured dotted lines. It is then natural to further cut the picture by drawing the gray dotted lines. In the right panel, we represent the eight hexagons arising from the procedure.
}\label{fig:toruswithdetails}
\end{center}
\end{figure}

As we have introduced eight hexagons, for both set of rapidities $\{u_1,u_2\}$ and $\{u_3, u_4\}$ we need an eight-fold partition, which we indicate as
$\underline{\alpha}$ and $\underline{\beta}$, respectively; here $\underline{\alpha} = \{\alpha_1,\ldots,\alpha_8\}$ with $\bigcup_{j=1}^8\alpha_j=\{u_1,u_2\}$, and similarly for $\beta$. In figure \ref{fig:toruswithdetails} we detail how to distribute such partitions. The partition factors $\omega$ are constructed as usual, \textit{cf.}\ eq.~\eqref{eq:BKVformula}. Somewhat schematically, we find
\begin{equation}
\begin{aligned}
\cA^8_{\ell_A \, \ell_C \, \ell_E \, \ell_G}  = \frac{L}{n\,{\cal G} \, S_{12}}  \sum_{\underline{\alpha}, \underline{\beta}}\quad &\Big\{\,\omega(\underline{\alpha},\underline{\ell}_\alpha)\,\omega(\underline{\beta},\underline{\ell}_{\beta})\\
& \quad\Big[\ \,\hhb{1}{3}{2}{\alpha_1}{}{\beta_2}_1 \
 \hhc{1}{2}{4}{\alpha_2}{\beta_1}{}_2  \\
 & \quad \quad\hhb{1}{4}{2}{\alpha_3}{}{\beta_4}_3 \
 \hhc{1}{2}{3}{\alpha_4}{\beta_3}{}_4  \\
 & \quad\quad\hhb{1}{3}{2}{\alpha_5}{}{\beta_6}_5 \
 \hhc{1}{2}{4}{\alpha_6}{\beta_5}{}_6  \\
 &\quad \quad\hhb{1}{4}{2}{\alpha_7}{}{\beta_8}_7 \
 \hhc{1}{2}{3}{\alpha_8}{\beta_7}{}_8\ \Big]\ \Big\},
 \end{aligned}
\end{equation}
where we have indexed the hexagons with a subscript corresponding to the labelling of $\alpha$-partitions in figure~\ref{fig:toruswithdetails}. Notice the combinatorial factor of $1/n$, where $n$ is the length of the longest cycle in the string $(\ell_A \, \ell_C \, \ell_E \, \ell_G)$. This avoids overcounting configurations due to cyclic symmetry; consider for instance $\cA^8_{1111}$. In this case the sum over partitions yields four identical terms, so that $n=4$; similarly, for \textit{e.g.}\ $\cA^8_{2121}$, $n=2$. The double sum over partitions might seem daunting at first.
Fortunately, the hexagon operator is constant whenever it contains  zero or one magnon; hence in the above formula only at most two $\widehat{\mathfrak{h}}$ factors can give a non-trivial dependence.

The amplitude $\cA^8_{ijkl}$ has almost the same symmetry properties as the colour factors $T_{ijkl}$: with one, two or three non-vanishing edge-widths the amplitude is totally symmetric under the exchange of the labels. Furthermore, all amplitudes with only two non-vanishing edge widths are equal to the $\cA^8_{\ell 000}$ case, which in turn coincides with the usual Gaudin norm on the sphere.
One difference is that, while the colour factor $T_{ijkl}$ is invariant under $S_4$ permutations, the hexagon amplitude $\cA^8_{ijkl}$ is not; rather, it is invariant under $\mathbb{Z}_4$ cyclic permutations, as for instance $ \cA^8_{2211} \neq  \cA^8_{2121}$; this ensures self-consistency of the definition of the cycle length~$n$.
Finally, we emphasise again that we only sum over genuine torus diagrams, \textit{i.e.}\ those where at least three edge-widths are non-zero. Note that, due to the presence of the colour factors, this has leading order $1/N^2$. We have listed in table~\ref{tab:torus} the torus contribution to the norm the of the first few BMN operators---those listed in table~\ref{tab:adim} in section~\ref{sec:treelvl}. Again, we find perfect matching between the field-theory construction and the integrability one.

\begin{table}[t]
\begin{center}
\begin{tabular}{l||r|c|l}
correlator & Field theory & $T_{ijkl}$ & hexagon amplitude \\
\hline
$\la \cB_4 \, \cB_4 \ra \; N^4 a_{12}^4$  & $ -2 * T_{2110} $ & $- N^2 + \ldots $  & $\cA^8_{2110} = -2$  \\
 & $+1 * T_{1111}$ & $+ N^2 + \ldots $ & $\cA^8_{1111} = +1$  \\
\hline
$\la \cB_5 \, \cB_5 \ra \; N^5 a_{12}^4$ & $+1 * T_{3110}$ & $- N^3 + \ldots $ & $\cA^8_{3110} = +1$  \\
& $ +1 * T_{2111} $ & $+ N^3 + \ldots$ & $\cA^8_{2111} = +1$  \\
\hline
& $+(1 \pm\sqrt{5}) * T_{4110}$ & $-N^4 + \ldots$ & $\cA^8_{4110} = +(1 \pm \sqrt{5})$ \\
& $+(3 \mp \sqrt{5}) * T_{3111}$ & $+N^4 + \ldots$ & $\cA^8_{3111} = +(3 \mp \sqrt{5})$  \\
$\la \cB_6^\mp \, \cB_6^\mp \ra \; N^6 \, a_{12}^4$ & $+\frac{1}{2}(1 \mp \sqrt{5}) * T_{2220}$ & $+N^4 + \ldots$ & $\cA^8_{2220} = +\frac{1}{2}(1 \mp \sqrt{5})$ \\
& $-\frac{1}{2}(1 \mp \sqrt{5}) * T_{2211}$ & $+N^4 + \ldots$ & $\cA^8_{2211} = -\frac{1}{2}(1 \mp \sqrt{5})$  \\
& $ +1 * T_{2121}$ & $+N^4 + \ldots$ & $\cA^8_{2121} = +1$  \\
\hline
& $ +5 * T_{5110}$ & $-N^5 + \ldots$ & $\cA^8_{5110} = +5$  \\
& $ +2 * T_{4111}$ & $+N^5 + \ldots$ & $\cA^8_{4111} = +2$  \\
$\la \cB_7' \, \cB_7' \ra \; N^7 a_{12}^4$ & $-1 * T_{3220}$ & $+N^5 + \ldots$ & $\cA^8_{3220} = -1$  \\
& $ +1 * T_{3211}$ & $+N^5 + \ldots$ & $2*\cA^8_{3211} = +1$  \\
& $ +2 * T_{3121}$ & $+N^5 + \ldots$ & $\cA^8_{3121} = +2$  \\
& $ + \frac{3}{2} * T_{2221}$ & $+N^5 + \ldots$ & $\cA^8_{2221} = +\frac{3}{2}$  \\
\hline
& $ +1 * T_{5110}$ & $-N^5 + \ldots$ & $\cA^8_{5110} = +1$  \\
& $ +4 * T_{4111}$ & $+N^5 + \ldots$ & $\cA^8_{4111} = +4$  \\
$\la \cB_7'' \, \cB_7'' \ra \; N^7 a_{12}^4$ & $+1 * T_{3220}$ & $+N^5 + \ldots$ & $\cA^8_{3220} = +1$ \\
& $ -1 * T_{3211}$ & $+N^5 + \ldots$ & $2*\cA^8_{3211} = -1$  \\
& $ +4 * T_{3121}$ & $+N^5 + \ldots$ & $\cA^8_{3121} = +4$  \\
& $ + \frac{5}{2} * T_{2221}$ & $+N^5 + \ldots$ & $\cA^8_{2221} = +\frac{5}{2}$
\end{tabular}
\end{center}
\caption{The torus part of two-point functions of single-trace operators.
For the two-point functions in the first column, we first list the result of Wick contractions for each given colour structure $T_{ijkl}$, \textit{cf.}\ eq.~\eqref{eq:Tcolour}. We also write down the leading-order term for the $1/N$ expansion of $T_{ijkl}$. Notice that we have rescaled the correlators to make more natural the field-theory $N$-counting; all the torus contributions we consider are $O(N^{L-2})$. We do not write sphere contributions from $T_{L000}$, which are of order $N^L$, and subleading contributions such as $T_{ij10}$, $i,j > 1$, see appendix~\ref{app:colour}.
The last column is the hexagon amplitude, and it matches field theory. Notice that in two cases there happens to be more than one way to embed one graph on the hexagon, similarly to what happened for four-point functions; we highlight this by writing \textit{e.g.}\ $2*\cA^8_{3211}$.
}
\label{tab:torus}
\end{table}

\subsection{Single-trace--double-trace correlators}
\label{sec:singletrdoubletr}

As we described around eq.~\eqref{eq:twoptnonplanar}, part of the $O(1/N^2)$ result comes from the two-point function of the single-trace part of each BMN operator with its double-trace admixtures. These contributions can also be calculated by hexagons. In particular, let us consider the four-hexagon tessellation of topology (42) in  figure~\ref{fig:fourtessellations}. We want to compute the overlap between a single trace operator $\cB_{L}$ and a double-trace operator $\cB_{L'}\cO_{L''}$. Firstly, notice that this correlator can be represented on a tessellation of topology (42) with $\cB_{L}$ at position $a_1$, $\cB_{L'}$ in position $a_2$ and $\cO_{L''}$ in position $a_4$, \textit{cf.}\ figure~\ref{fig:fourtessellations}. By doing this, we implicitly introduce a point-splitting regularisation. However, notice that $\cO_{L''}$ and $\cB_{L'}$ are always placed on distinct hexagons. Therefore, we can safely and straightforwardly take the limit $a_4\to a_2$ in our result.

The hexagon amplitude therefore gives
\begin{equation}
\langle\cB_L; \cB_{L'} \cO_{L''}\rangle = \frac{\sqrt{L\,L'\,L''}}{\cG_{L} \cG_{L'} S_{12} S_{1'2'}}  \, \cA^{(42)}_{00L'} \, ,
\end{equation}
where the S matrices $S_{12}$ and $S_{1'2'}$ scatter the two magnons on $\cB_{L}$ and $\cB_{L'}$, respectively.
The color dressing of the hexagon in this case is trivial, as the leading-order term is universal and equal to $1/N$. We find
\begin{equation}
\begin{gathered}
\la \cB_6\mp; \cB_4 \cO_2 \ra  = \frac{7 \mp \sqrt{5}}{\sqrt{6} \, N \, a_{12}^4} + \ldots \,, \qquad
\la \cB_7'; \cB_4 \cO_3 \ra  =  \frac{\sqrt{2}}{N \, a_{12}^4} +\ldots \, ,\\
\la \cB_7'; \cB_5 \cO_2 \ra  =  \frac{5}{\sqrt{2} \, N \, a_{12}^4} + \ldots \, ,  \qquad
\la \cB_7''; \cB_4 \cO_3 \ra  =  - \frac{2 \sqrt{6}}{N \, a_{12}^4} + \ldots \, ,\\
\la \cB_7''; \cB_5 \cO_2 \ra  =  - \frac{3 \sqrt{3}}{\sqrt{2} \, N \, a_{12}^4} + \ldots \, ,
\end{gathered}
\end{equation}
in full agreement with free-field theory.

For the admixtures of operator $\cB_7'$  of equation \eqref{eq:B7primeadmixt}, we therefore find a contribution for the single-trace--double-trace overlap which reads
\begin{equation}
2 \, \left\la \cB_7' \Big( - \frac{3}{2 \sqrt{2} N} \cO_3 \cB_4 - \frac{2 \sqrt{2}}{N} \cO_2 \cB_5 \Big) \right\ra = - \frac{23}{N^2 a_{12}^4} + \ldots\,.
\end{equation}
The only remaining contribution we need in order to find the $1/N^2$ order of eq.~\eqref{eq:twoptnonplanar} is the double-trace--double-trace term, which is dominated by the disconnected contribution. This can be easily found, and reads
\begin{equation}
\frac{9}{8 \, N^2} \la \cO_3 \cO_3 \ra \la \cB_4 \cB_4 \ra + \frac{8}{N^2} \la \cO_2 \cO_2 \ra \la \cB_5 \cB_5 \ra + \ldots = \frac{73}{8} \frac{1}{N^2 a_{12}^4} +\ldots \, .
\end{equation}

\section{Hexagons, L\"uscher corrections and Feynman graphs}
\label{sec:Nis2}

We have seen that colour-ordering is necessary to correctly reproduce generic four-point functions as well as the torus part of the norm. We need this prescription both to exclude certain tree-level graphs which are sub-leading (or vanishing) in the $1/N$ expansion, as well as to account for non-trivial minus signs in other graphs. Based on this experience, it is natural to wonder whether the empirical rule to exclude the wrapping contribution of ``edge reducible'' graphs proposed in ref.~\cite{Fleury:2016ykk} (see also section~\ref{sec:edgered}) might also be understood in terms of colour factors. In the section below we show that this is indeed the case, at least at one loop and for $\tfrac{1}{2}$-BPS operators. We will also highlight a rather direct link between $\cN =2$ Feynman graphs and the L\"uscher-like corrections which encode finite-size effects in the hexagon formalism.

To begin with, we briefly review how to rephrase $\cN = 4$ SYM in terms of  $\cN = 1$ and $\cN = 2$ supermultiplets, which will allow us to formulate the Drukker-Plefka kinematics~\cite{Drukker:2009sf} in terms $\cN = 2$ multiplets.

\subsection{\texorpdfstring{$\cN = 2$}{N=2} superfields for \texorpdfstring{$\cN = 4$}{N=4} SYM}

Here we give a brief account at the linearised level of how the components of the $\cN=4$ field-strength multiplet can be arranged into $\cN=1$ and $\cN=2$ multiplets  in the Wess-Zumino gauge. To obtain an off-shell quantum formalism the multiplets have to be enlarged in both cases by further components (``subcanonical'' and ``auxiliary'' fields). For a full account of the superfield formulations we refer the reader to ref.~\cite{Wess:1992cp} for $\cN=1$ diagrams and to ref.~\cite{Galperin:2001uw} for $\cN=2$ supergraphs.

The list of elementary fields of the $\cN = 4$ model comprises three complex scalars, four Majorana-Weyl fermions, and the field strength of the gauge potential $A_\mu$:
\begin{equation}
F_{\mu \nu} \, , \qquad  \psi^I_\alpha, \, \bar \psi_{I \dot \alpha} \,  \qquad \phi^{[IJ]} \, , \  \ \phi_{[KL]} = \overline{\phi^{[KL]}} = \frac{1}{2} \epsilon_{KLIJ} \phi^{[IJ]} \, , \qquad I,J,K,L=1,\dots4
\end{equation}
All of these transform in the adjoint representation of a non-abelian gauge group; integrability arises in the case of $SU(N)$. Introducing Grassmann parameters $\theta^I_\alpha, \, \bar \theta_{I \dot \alpha}$ for the on-shell supersymmetry of the multiplet we might try to construct a superfield
\begin{equation}
\varphi^{[IJ]} = \phi^{[IJ]} + \theta^{[I}_\alpha \psi^{J] \alpha} + \frac{1}{2}  \epsilon^{IJKL} \bar \theta_{K \dot \alpha} \bar \psi_L^{\dot \alpha} + \theta^{[I}_\alpha \theta^{J]}_\beta \, F^{\alpha \beta} + \frac{1}{2} \epsilon^{IJKL} \bar \theta_{K \dot \alpha} \bar \theta_{L \dot \beta} F^{\dot \alpha \dot \beta} \, .
\end{equation}
Unfortunately, till now there is no superspace formulation that makes the entire $\cN = 4$ supersymmetry manifest (\textit{i.e.}, that realises it on the coordinates of an extension of Minkowski space) because the supersymmetry transformations close only on shell, \textit{i.e.}\ up to equations of motion. For subsets of the supersymmetry generators this goal can be achieved, though.

One frequently-employed approach is to keep only  $\theta^1, \bar \theta_1$. Then $\varphi^{IJ}$ from the last equation breaks into
\begin{equation}
\Phi^i = \varphi^{1i} = \phi^{1i} + \frac{1}{2} \theta^1 \psi^i \, , \qquad i \in \{2,3,4\} \, ,
\end{equation}
which gives three complex chiral fields (and their conjugates). The leftover components $\psi^1$ and $F_{\alpha\beta}$ are put into
\begin{equation}
W_\alpha = \psi^1_\alpha + \theta^{1 \beta} F_{\beta \alpha} \, ,
\end{equation}
and its conjugate---the $\cN = 1$ Yang-Mills multiplets. Introducing additional ``auxiliary'' fields  these multiplets can be extended to superfields with $\cN = 1$ off-shell supersymmetry~\cite{Wess:1992cp}.

Alternatively, we can keep $\theta^i, \bar \theta_i$ with $i=1,2$. This yields a complex doublet
\begin{equation}
q^i = \varphi^{i4} = \phi^{i4} + \frac{1}{2} \theta^i_\alpha \psi^{4 \alpha} + \frac{1}{2} \epsilon^{i4j3} \bar \theta_{j \dot \alpha} \bar \psi_3^{\dot \alpha} \, ,
\end{equation}
the ``hypermultiplet'', and a complex singlet
\begin{equation}
W = \varphi^{12} = \phi^{12} + \frac{1}{2} \theta^{i \alpha} \psi_{i \alpha} + \frac{1}{2} \theta^i_\alpha \theta_{i \beta} F^{\alpha \beta} \, ,
\end{equation}
the $\cN = 2$ Yang-Mills multiplet. In passing we have introduced an antisymmetric symbol $\epsilon_{ij}$ that can be used to lower and raise internal $i$ indices. Obviously, these fields are not real; rather, they are supplemented by their complex conjugates.

The problem of introducing auxiliary fields for the N=2 multiplets was
resolved in refs.~\cite{Galperin:1984av, Galperin:1985va,Galperin:1985bj} (see also the references therein w.r.t.\ alternative approaches) by resorting to "harmonic superspace", which has an additional bosonic variable $u^{\pm i} \in SU(2)/U(1)$.  Here the row index is written as $\pm$ to denote the charge under the $U(1)$ group in the coset. We will not need the details of the formalism since we will simply import the result we need from ref.~\cite{Howe:1999hz}. What we will exploit, though, is that the doublet $q^i$ and its complex conjugate $\bar q_i$ are both projected by the first row of the matrix $u$, yielding
\begin{equation}
q^+ = u^+_i \, q^i \, , \qquad \tilde q^+ = u^{+i} \, \bar q_i \, .
\end{equation}

Last, the field $W$ can be written as a superspace derivative of a pre-potential $V^{++}$ which is the second dynamical field in the formulation of ref.~\cite{Galperin:1984av,Galperin:1985va,Galperin:1985bj}. The $\cN = 4$ SYM action is then
\begin{equation}
S = -\int d^4x_A d^2\theta^- d^2 \bar \theta^- du \, \text{Tr}\big[q^+ (D^{++} \tilde q^+ + i\,[V^{++}, \tilde q^+])\big] + \frac{1}{4g^2} \int d^4x_L d^4 \theta \, \text{Tr}[W^2] \, .
\label{n2Action}
\end{equation}
Here, the coordinates $x_A$ are shifted by Grassmann parameters  with respect to the Min\-kowski ones~\cite{Howe:1999hz}; this is similar to what happens for coordinates $x_L$ in the chiral basis~\cite{Galperin:1984av,Galperin:1985va,Galperin:1985bj, Wess:1992cp}.
Notice that, even if the whole Lagrangian is fairly intricate due to the presence of the pre-potential~$V$, the matter sectors have simple interactions. This makes the $\cN=1$ and $\cN=2$ formulations useful for describing correlators that have only matter fields at the external points---\textit{i.e.}, that have chiral fields in $\cN=1$ or hypermultiplets in $\cN=2$ at external points. Let us focus on the latter case; observe that in $\cN=2$ the only relevant interaction is the cubic vertex $\text{Tr}\big[q\,[V,\tilde{q}]\big]$ at one loop. Generically, for correlators with external hypermultiplet fields, the $\cN = 2$ Feynman rules amount to decorating skeleton graphs with virtual particles---\textit{Yang-Mills (YM) lines}, which propagate the $V^{++}$ field from the Yang-Mills multiplet. This is similar in spirit to the what happens in the integrability picture, where one decorates a tree-level diagram by mirror particles~\cite{Basso:2015zoa,Fleury:2016ykk}. Below we make this correspondence explicit at one loop.

Finally, remark that the conformal invariance of $\cN = 4$ SYM is not manifest in individual diagrams. It only arises in the sum over graphs after skilful handling of numerator algebra~\cite{GonzalezRey:1998tk, Eden:1998hh, Eden:1999kh, Bianchi:1999ge}.
A convenient way to compute one-loop interactions in correlators of composite operators built out of hypermultiplets is to differentiate the path integral~\cite{Howe:1999hz}:
\begin{equation}
\frac{\partial}{\partial g^2} \la \cO_1 \ldots \cO_n \ra = \frac{-i}{4g^4} \int d^4x_{L0} d^4\theta_{0} \la \, \tr(W^2(x_0,\theta_0) \, \cO_1 \ldots \cO_n \ra\,.
\end{equation}
In particular, this directly yields the ``one-loop box'', \textit{i.e.}\ the only one-loop conformal integral. We refer the reader to the original paper ref.~\cite{Howe:1999hz}; here we will only make use of a result of that paper,  \textit{cf.}\ eq.~(\ref{eq:supergraph:f1234}) below.

\subsection{The Drukker-Plefka vacuum  as a sum of hypermultiplets}

We have seen in section~\ref{sec:treelvl} that the Drukker-Pleka vacuum $\text{Tr}[\mathcal{Z}^L]$ can be parametrised as in eq.~\eqref{eq:drukkerplefka}. The R-symmetry coordinate dependence can be written in terms of the six scalars of $\cN=4$ SYM, $\Phi=(\varphi^i)_{i=1,\dots 6}$ and of a vector $\eta$,
\begin{equation}
\mathcal{Z} = \eta \cdot \Phi \, , \qquad \eta = \left( \frac{1 + \alpha \bar \alpha}{2}, \, i \frac{1 - \alpha \bar \alpha}{2}, \, i \, \text{Im}\,{\alpha}, \, i \, \text{Re}\,{\alpha}, \, 0, \, 0 \right)  .
\label{eq:so6param:vacuum}
\end{equation}
We reproduce this in the $\cN=2$ language by assembling the complex scalars $Z,Y$ into hypermultiplets. This can be done in two ways: either
\begin{equation}
\label{eq:Nis2param:vacuum}
q^i = (Z,Y),  \quad
\bar{q}_i = (\bar{Z}, \bar{Y}),
\quad u_i^+ = (1,\alpha),
\qquad
\mathcal{Z} = q^+ + \bar{\alpha} \, \tilde q^+\,,
\end{equation}
or alternatively,
\begin{equation}
q^i = (Z,\bar Y), \quad
\bar{q}_i = (\bar Z, Y),\quad
u_i^+ = (1,-\bar \alpha),
\qquad
\mathcal{Z} = q^+ + \alpha \, \tilde q^+ \, .
\end{equation}
We will adopt the first choice, but it is clear from the existence of the second scheme that $\cN = 4$ results will have to be $\alpha \leftrightarrow \bar \alpha$ symmetric. Notice that $(1,\alpha)$ can be completed to an element $u$ of $SU(2)/U(1)$,
\begin{equation}
u = \left(
\begin{array}{c}
u^+ \\ u^-
\end{array}
\right)
= \frac{1}{\sqrt{1 + \alpha \bar \alpha}}
\left(
\begin{array}{cc}
1 & \alpha \\ -\bar \alpha & 1
\end{array}
\right)\,.
\end{equation}
The normalisation factor $\sqrt{1 + \alpha \bar \alpha}$ is irrelevant in what follows, as our formula eq.~(\ref{eq:supergraph:f1234}) below is homogeneous in~$u$.
Finally, notice that in the above construction we did not consider the ``transverse excitation'' $X$ and its conjugate $\bar{X}$. In fact, $X$ is the lowest component of the $\cN=2$ YM multiplet.

\subsection{One-loop diagrams}

\begin{figure}[t]
\begin{center}
\includegraphics[height = 25 mm]{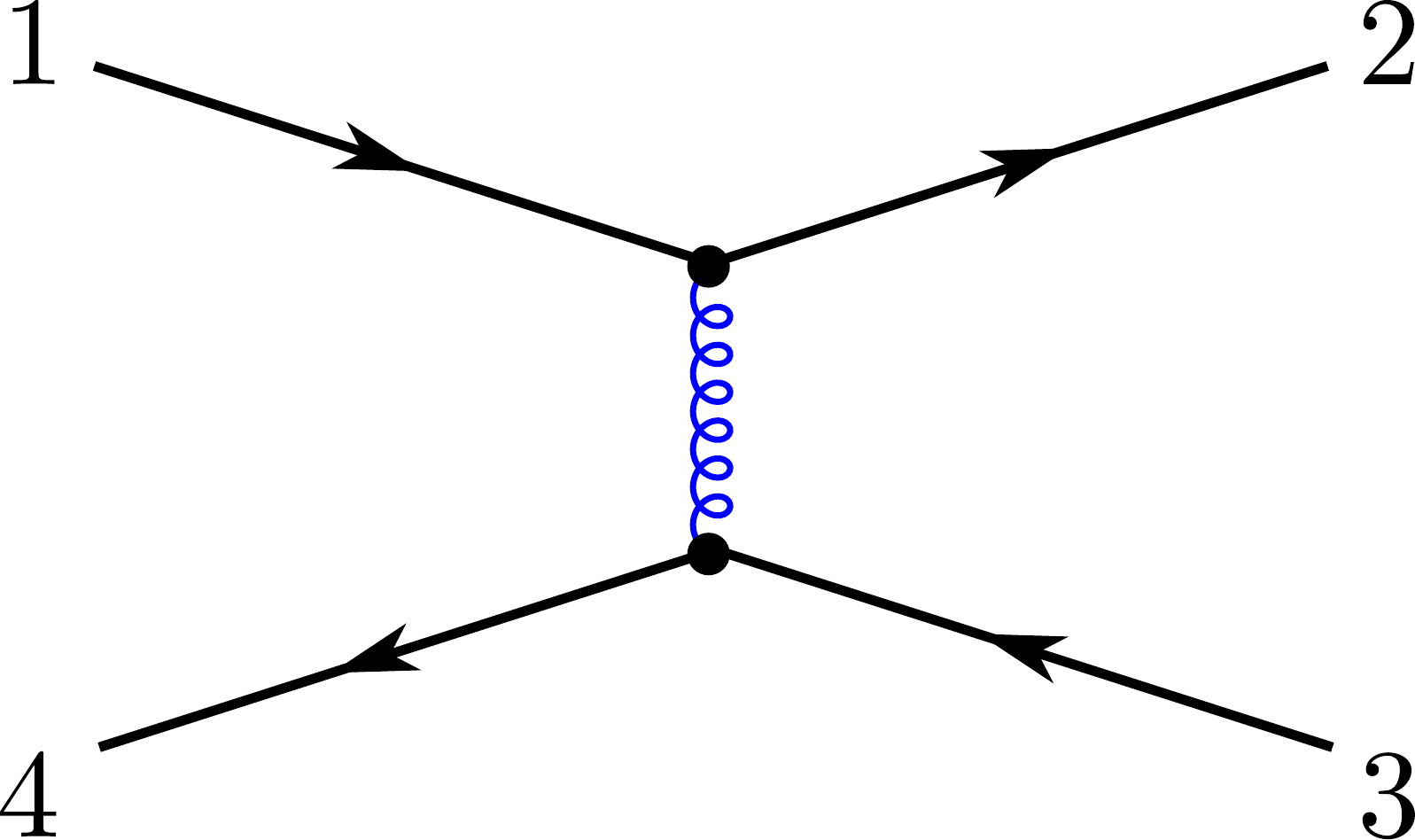}
\caption{We depict the exchange of a $\cN = 2$ Yang-Mills multiplet as a wavy line; the straight, directed lines are hypermultiplet propagators.}
\label{fig:YMexchange}
\end{center}
\end{figure}

We now want to consider four-point functions of $\tfrac{1}{2}$-BPS operators $\cO_L = \text{Tr}[\mathcal{Z}^L]$. At tree level (and $\theta, \bar \theta = 0$) the graphs are simply of products of hypermultiplet propagators%
\footnote{We introduce the intuitive short-hand notation $\tilde{q}_1$ to indicate that the fields are at position $x_1$ in Minkowski space and have internal coordinate $\alpha_1$, and so on.}
\begin{equation}
\langle \tilde q_1 \, q_2 \rangle = \frac{(12)}{x_{12}^2} \, , \qquad (12) = u_1^{i+} u^+_{2i} =\alpha_1-\alpha_2\, .
\end{equation}
This propagator is antisymmetric under the point exchange $1 \leftrightarrow 2$ because of its numerator. At one loop, a single Yang-Mills line is inserted in all possible ways into the tree graphs. We will show that this is exactly equivalent to the exchange of virtual magnons in the integrability picture.

The simplest diagram we need to compute, which will be the building block for the rest of our analysis, is thus given by two hypermultiplet lines between points (12) and (34), connected by a Yang-Mills exchange, \textit{cf.}\ figure~\ref{fig:YMexchange}.
Evaluating this supergraph by a Lagrangian insertion~\cite{Howe:1999hz} yields:
\begin{equation}
f_{12;34} = T_{12;34} \left[ \frac{(12)}{x_{12}^2} \frac{(34)}{x_{34}^2} (x_{14}^2 x_{23}^2 - x_{13}^2 x_{24}^2) + (13)(24) + (14)(23) \right] \, g_{1234} + \ldots\,,
\label{eq:supergraph:f1234}
\end{equation}
with the colour factor
\begin{equation}
T_{12;34} = \text{Tr}([T_1,T_2] [T_3,T_4]),
\end{equation}
which is a double commutator of the gauge group generators $T_1, \ldots,T_4$ in the adjoint representation carried by the hypermultiplets at the outer points.
Moreover, in eq.~(\ref{eq:supergraph:f1234}) we have the finite and conformal one-loop box integral
\begin{equation}
g_{1234} = \int \frac{d^4x_5}{x_{15}^2 x_{25}^2 x_{35}^2 x_{45}^2} \, .
\end{equation}
Finally, the ellipsis in eq.~\eqref{eq:supergraph:f1234} indicates three-point and two-point integrals that must (and will) cancel in complete BPS correlators due to conformal invariance. 
Notice that the one-loop box~$g_{1234}$ is fully symmetric under point exchange, while the rational expression in the square brackets in (\ref{eq:supergraph:f1234}) is symmetric under both $1 \leftrightarrow 2$ and $3 \leftrightarrow 4$ separately. Due to the colour factor, the complete block $f_{12;34}$ works out to be antisymmetric under these exchanges. Hence it has the same point-flip properties as the corresponding tree-level graph---a pair of free hypermultiplet propagators without any YM line.

Next, with the assignment (\ref{eq:Nis2param:vacuum}) for $\mathcal{Z}$ we find
\begin{equation}
\langle \mathcal{Z}_1 \, \mathcal{Z}_2 \rangle = \bar \alpha_1 \frac{(12)}{x_{12}^2} + \bar \alpha_2 \frac{(21)}{x_{21}^2} = \frac{\alpha_{12} \bar \alpha_{12}}{x_{12}^2} = \frac{y_{12}^2}{x_{12}^2} \, ,
\end{equation}
where $\alpha_{ij}=\alpha_i-\alpha_j$, $y_{ij}=y_i-y_j$, and the four-vector $y_i$ is given by the last four components of the six-vector $\eta$ in eq.~(\ref{eq:so6param:vacuum}) evaluated at $\alpha_i$.
The r.h.s.\ $y^2/x^2$ is the correct free-field theory two-point function for an $\cN=4$ field strength multiplet; indeed, even if our formalism only explicitly preserved $\cN=2$ supersymmetry off-shell, our final result is compatible with full on-shell supersymmetry as it must.

\begin{figure}[t]
\includegraphics[width=\linewidth]{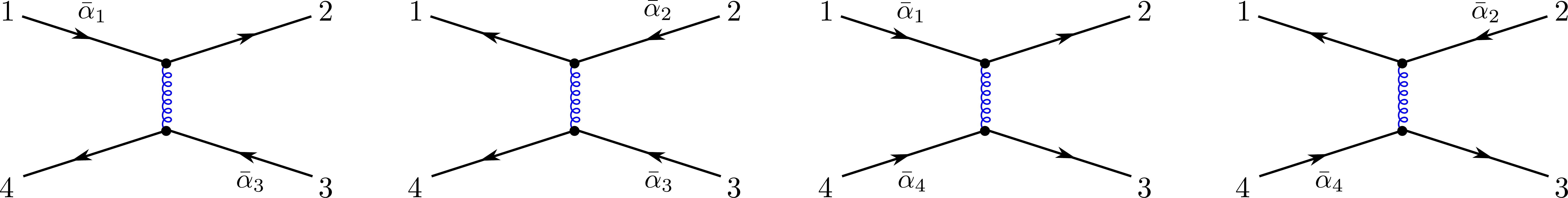}
\caption{We depict the four $\cN = 2$ YM exchanges which add up to the full one-loop exchange in $\cN = 4$ SYM.}
\label{fig:fourYMexchanges}
\end{figure}

Using eq.~(\ref{eq:supergraph:f1234}), we can compute the one-loop contribution to the graph with matter lines $\langle \mathcal{Z}_1 \, \mathcal{Z}_2 \rangle$ and $\langle \mathcal{Z}_3 \, \mathcal{Z}_4 \rangle$, see figure~\ref{fig:fourYMexchanges}. The four diagrams of that figure combine to give
\begin{equation}
F_{12;34} = \bar \alpha_{12} \bar \alpha_{34} \, f_{12;34}\,,
\end{equation}
due to the antisymmetry of $f_{12;34}$ under $1 \leftrightarrow 2$ and $3 \leftrightarrow 4$. The expression we found for $F_{12;34}$ cannot be written in terms of $y^2/x^2$ propagators. By elementary manipulations we can recast it as
\begin{equation}
F_{12;34} = T_{12;34} \left[ \frac{y_{12}^2}{x_{12}^2} \frac{y_{34}^2}{x_{34}^2} (x_{14}^2 x_{23}^2 - x_{13}^2 x_{24}^2) + y_{13}^2 y_{24}^2 - y_{14}^2 y_{23}^2 \, + \,w \right] \,  g_{1234} \, , \label{n4F}
\end{equation}
where
\begin{equation}
w = \bar{\alpha}_{13}\,\bar{\alpha}_{24}\,\alpha_{14}\,\alpha_{23}
-\bar{\alpha}_{14}\,\bar{\alpha}_{23}\,\alpha_{13}\,\alpha_{24}\,.
\end{equation}
Notice that the function $w$ is still not expressed in terms of $y_{ij}^2$; moreover, it is antisymmetric under $\alpha \leftrightarrow \bar \alpha$, rather than symmetric as expected.
Indeed if we take into account that the field $\mathcal{Z}$ contains both $q$ and $\tilde{q}$ and sum over all diagrams, we always find the combination $F_{12;34} + F_{13;24} + F_{14;23}$. Using the Jacobi identity $T_{12;34} - T_{13;24} + T_{14;23}=0$, we conclude that $w$ cancels in the final result due to its antisymmetry.
Hence we can write our result for $F_{12;34}$ as
\begin{equation}
F_{12;34} = T_{12;34} \, \Pi_{12} \Pi_{34} \, g_{1234} \ \tilde F_{12;34}\,, \label{tildeF}
\end{equation}
where
\begin{equation}\
\Pi_{ij}=\pp{ij}\,,\qquad	
\tilde F_{12;34} = x_{14}^2 x_{23}^2 - x_{13}^2 x_{24}^2 - x_{12}^2 x_{34}^2 \left(\frac{y_{14}^2 y_{23}^2}{y_{12}^2 y_{34}^2} - \frac{y_{13}^2 y_{24}^2}{y_{12}^2 y_{34}^2} \right) .
\end{equation}
This seems like an odd choice---we could have multiplied in the factor $x_{12}^{-2} x_{34}^{-2}$ and manifestly obtained a function of the cross ratios. However, this splitting highlights that $F_{12;34}$ corresponds to inserting a line inside a tree-level graph with propagators $\Pi_{12} \Pi_{34}$.

\begin{figure}[t]
\begin{center}
\includegraphics[width=\linewidth]{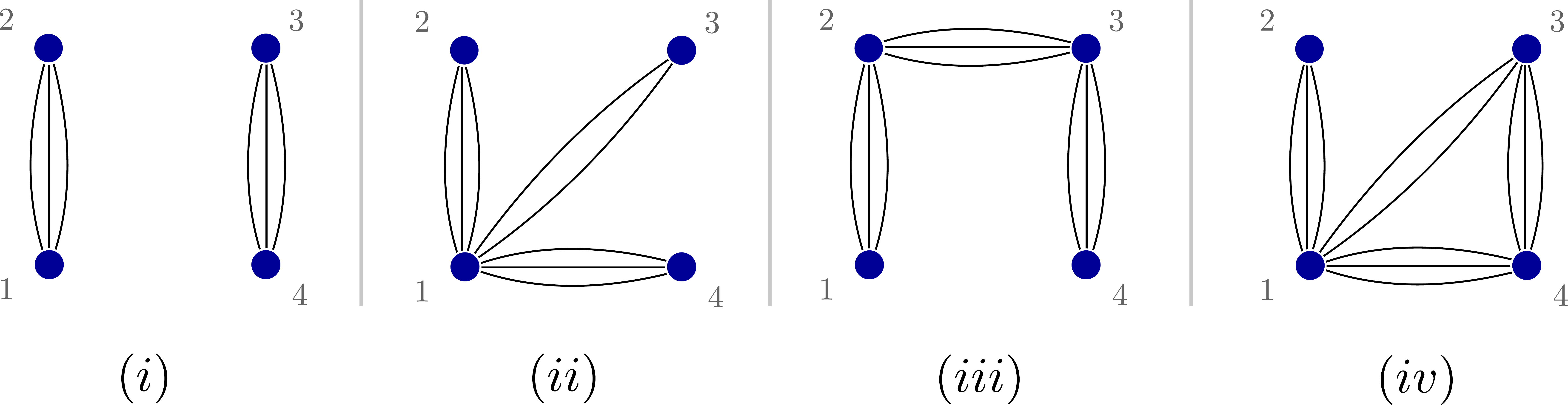}
\caption{Four classes of one-edge reducible graphs; each graph can be disconnected by cutting a single edge with however many propagators.
}
\label{fig:edgered}
\end{center}
\end{figure}

\subsection{Edge-reducible graphs}
Let us now evaluate the corrections for the ``edge-reducible'' graphs of section~\ref{sec:edgered}. We can group them in four categories, which we represent in figure~\ref{fig:edgered}. In terms of the propagators $\Pi_{ij}=y_{ij}^2/x_{ij}^2$, at tree level we have%
\footnote{%
We call the graphs of type $(iv)$ ``subextremal'' because they have the same topology as the ones we studied in section~\ref{sec:subextremal}, possibly with different bridge-lengths.
}
\begin{equation}
\begin{aligned}
(i) &\quad \Pi_{12}^{\ell_{12}} \, \Pi_{34}^{\ell_{34}} &&\quad \text{disconnected} \, , \\
(ii) &\quad \Pi_{12}^{\ell_{12}} \, \Pi_{13}^{\ell_{13}} \, \Pi_{14}^{\ell_{14}} &&\quad \text{extremal} \, ,\\
(iii) &\quad \Pi_{12}^{\ell_{12}} \, \Pi_{23}^{\ell_{23}} \, \Pi_{34}^{\ell_{34}} &&\quad \text{sausages} \, ,\\
(iv) &\quad \Pi_{12}^{\ell_{12}} \, \Pi_{13}^{\ell_{13}} \, \Pi_{14}^{\ell_{14}} \, \Pi_{34}^{\ell_{34}} &&\quad \text{subextremal}  \, .
\end{aligned}
\end{equation}
These graphs can be disconnected by cutting one edge. The claim of ref.~\cite{Fleury:2016ykk} is that contributions of mirror magnons due to these graphs should not be included in the hexagon formalism. Below, we argue that at one-loop all the colour factors of these diagrams vanish.

\begin{figure}[t]
\begin{center}
\includegraphics[width=\linewidth]{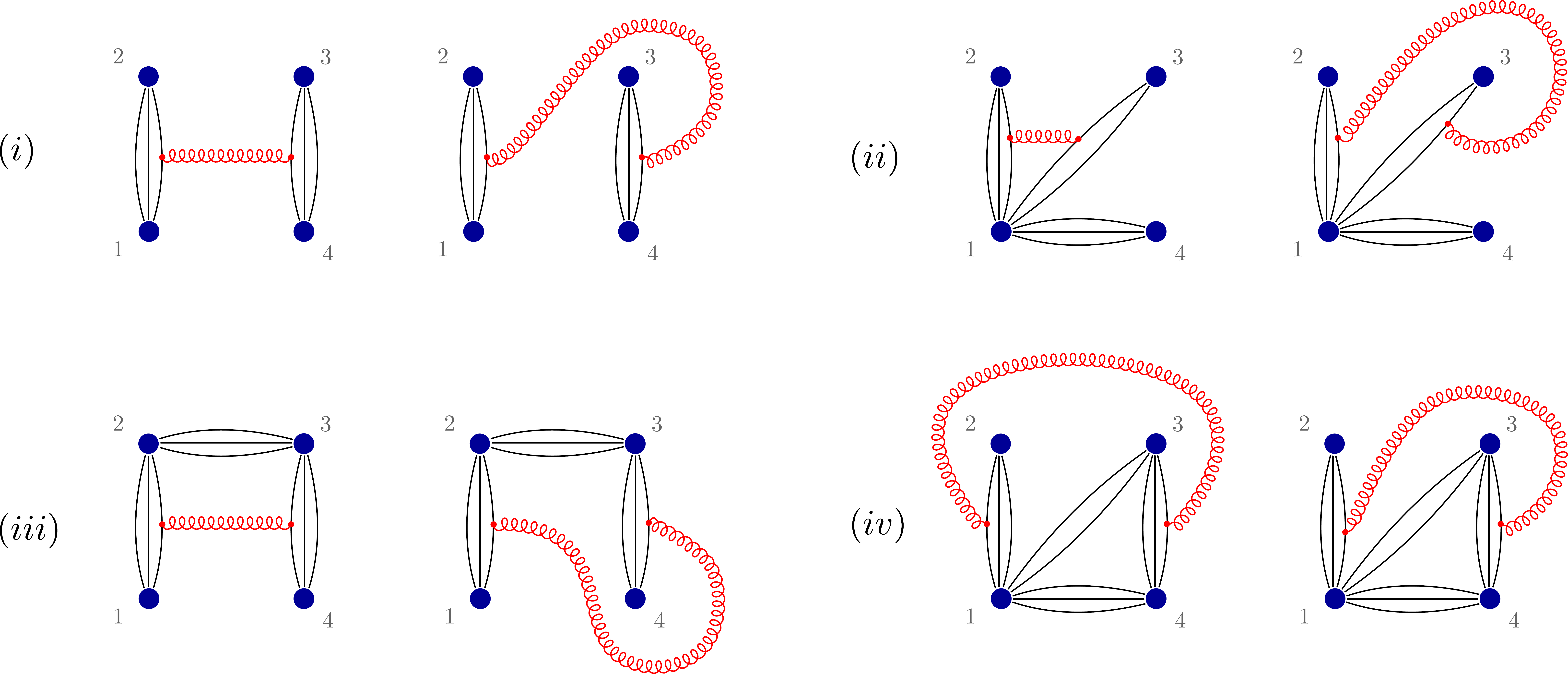}
\caption{Structure of the one-loop diagrams for the four ``edge-reducible'' graphs. We planarly attach a YM line to the matter propagators in all possible ways. As described in the text, the colour factor of $(i)$ is zero for each diagram. For $(ii)$, there is no contribution to the four-point integral, since only three-point topologies arise. For case $(iii)$ the colour factors only cancel between the two diagrams depicted, and similarly for case~$(iv)$.
}
\label{fig:oneloopgraphs}
\end{center}
\end{figure}

\begin{figure}[t]
\begin{center}
\includegraphics[width=0.9\linewidth]{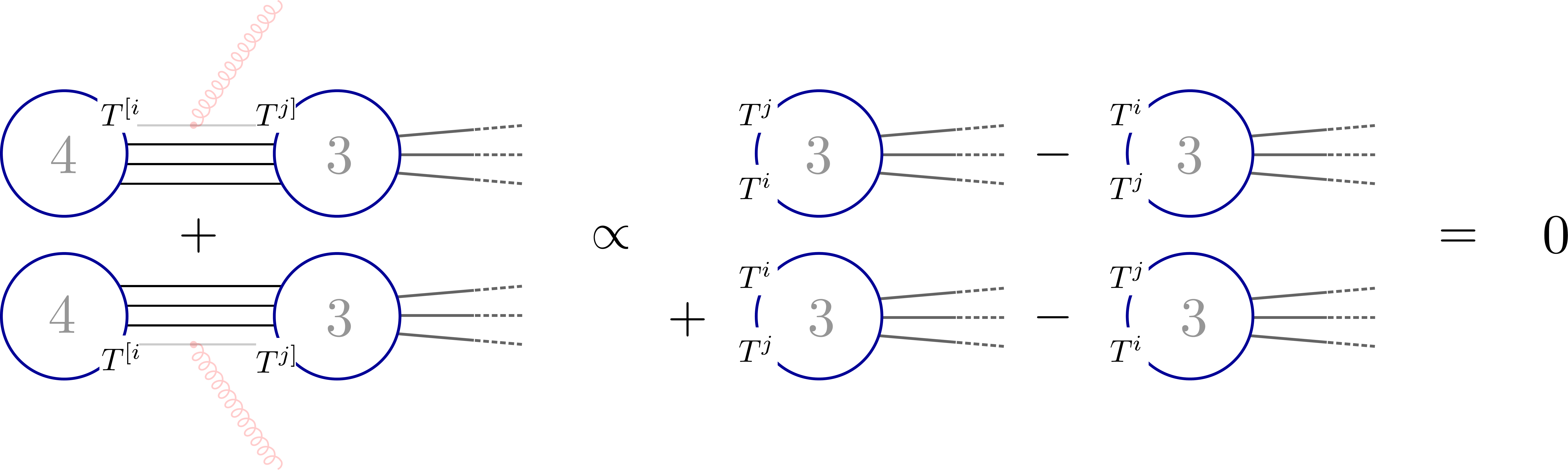}
\caption{We sketch the cancellation of the leading large-$N$ term for diagrams of type~$(iii)$. We start on the top left diagram, which corresponds to part of the first graph in panel $(iii)$ of figure~\ref{fig:oneloopgraphs}.
Looking at operator~$4$, we have a number of matter lines to which colour generators are attached; to the first of those lines, a YM line is attached. The vertex carries a commutator of the colour generators. Below, we have a similar structure, with the exception that the YM line is now attached to the \textit{bottom} matter line; this corresponds to the second diagram in panel $(iii)$ of figure~\ref{fig:oneloopgraphs}. By using the colour-trace identities of appendix~\ref{app:colour}, we can eliminate the trace corresponding to operator four. We are left with four contributions that cancel each other.
}
\label{fig:colourtraces}
\end{center}
\end{figure}

This is obvious for the disconnected graphs, \textit{i.e.}\ for case $(i)$. The YM line can only be inserted between the two strands of (13) and (24) propagators as in figure~\ref{fig:oneloopgraphs} if we want to find the four-point block~$F_{13;24}$. Then at one loop the colour factor vanishes much like it was the case for the one-particle reducible graph discussed in section~\ref{sec:colourdressing}. Case $(ii)$ also does not contribute to the four-point function at this order. In fact, the YM line always gives the structure of a three-point function, see figure~\ref{fig:oneloopgraphs}. Such contributions cancel as required by conformal invariance.
The remaining cases $(iii)$ and $(iv)$ are more subtle. Let us start from case~$(iii)$. From the structure of the matter propagators, notice that the only expression of the type $F_{ij;kl}$ which may arise is $F_{12;34}$. There are two ways to attach a YM~line to the tree level graph, and they are depicted in figure~\ref{fig:oneloopgraphs}. As remarked, both graphs will be proportional to $F_{12;34}$, up to their colour factors. As we depict graphically in figure~\ref{fig:colourtraces}, the sum of the two colour factors actually vanishes.
Hence, for $(iii)$ planar one-loop quantum corrections cancel; the same happens in case $(iv)$  by a similar argument. Notice that this argument does not make use of any properties of the function~$F_{12;34}$.

\begin{figure}[t]
\begin{center}
\includegraphics[width=0.25\linewidth]{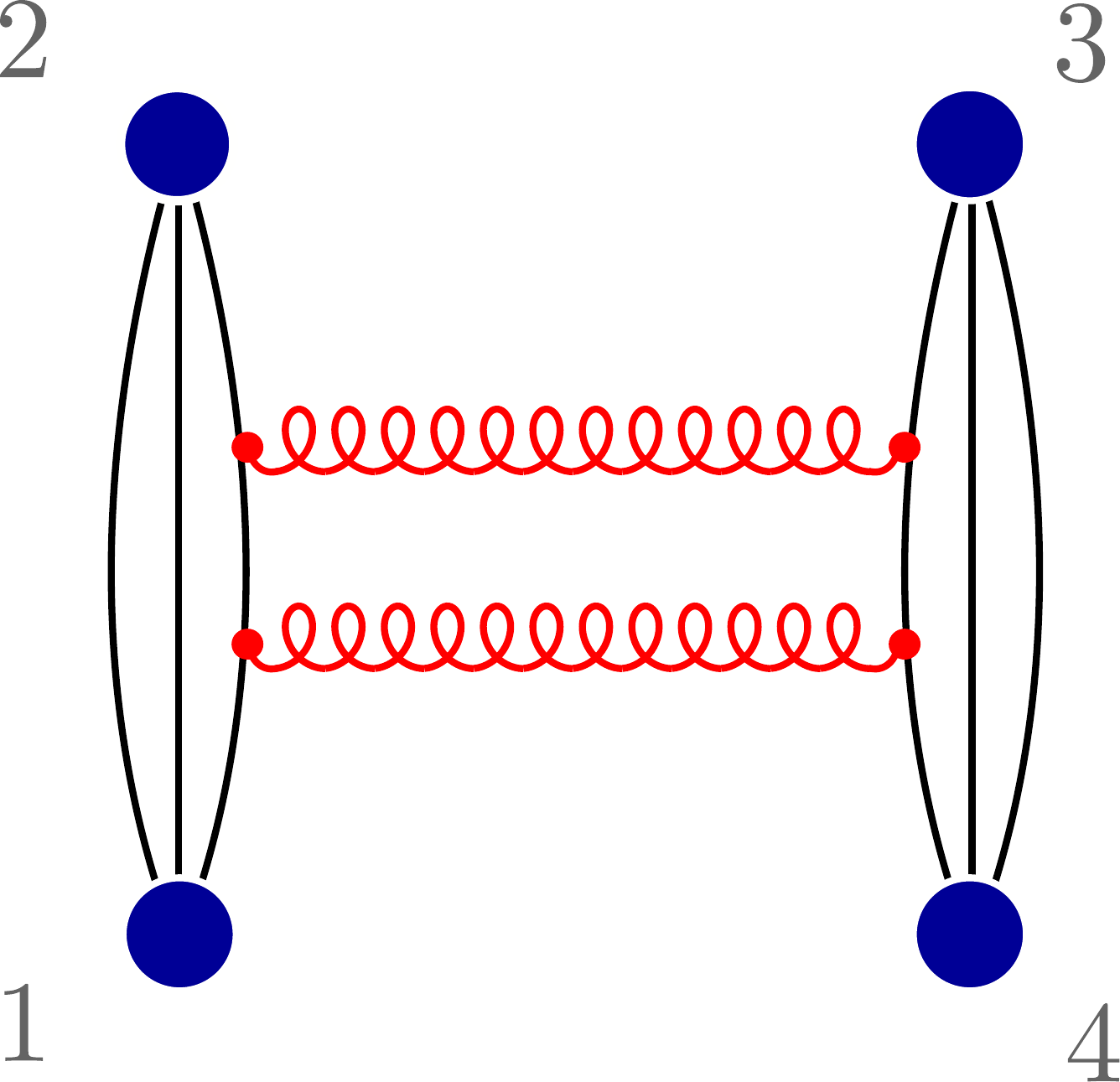}
\caption{At two loops, we expect some of the graphs which did not contribute at lower orders to appear, as they have a leading-order colour factor. One such example is the graph depicted, which is disconnected at tree-level.
}
\label{fig:twoloops}
\end{center}
\end{figure}

We conclude that, at one-loop, the ``edge-reducibility'' criterion of ref.~\cite{Fleury:2016ykk} is reproduced by our colour-dressing procedure for the case of four-point functions of $\tfrac{1}{2}$-BPS operators at one loop. In fact, since we never used the explicit form of $F_{12;34}$, it is easy to extend these arguments to the case where the operators contain $sl(2)$ excitations.
However, it is worth emphasising that the two criteria are different beyond one-loop: not only this is the case at tree level, as discussed at length above, but we also expect discrepancies at two~loops. For instance, if we decorate a disconnected graph by \textit{two} Yang-Mills lines like in figure~\ref{fig:twoloops}, we expect to find a contribution in field theory, \textit{cf.}\ ref.~\cite{Eden:2000mv}. It would be interesting to analyse the structure of two- and possibly higher-loops graph and compare it with the hexagon approach.

\subsection{Mirror magnons as Yang-Mills lines}
\label{sec:wrapping}
So far we have focussed on the diagrams that \textit{do not} contribute to the four point function of BPS operators. Let us now look at those that give non-zero contributions. One class of diagrams is the one where all six bridge-lengths are non-zero, \textit{cf.}\ figure~\ref{fig:fourtessellations}. These diagrams do contribute to wrapping effects, but only at two or more loops~\cite{Fleury:2016ykk}. Therefore, they will not be important in our discussion.

\begin{figure}[t]
\begin{center}
\includegraphics[width=\linewidth]{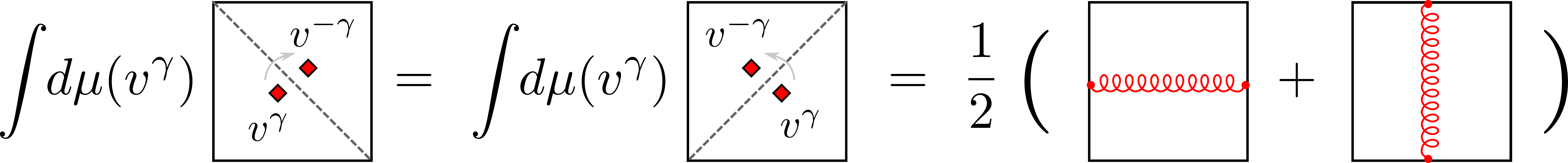}
\caption{At one-loop, gluing over a width-zero edge (one of the two diagonals of the square in the figure) is equal to the semi-sum of the two possible planar Yang-Mills exchanges in which the virtual particle crosses that diagonal.
}
\label{fig:hexagonFTmatch}
\end{center}
\end{figure}

We are left with four-point functions given by a square, or a square with one diagonal. Let us start from the former case---four non-vanishing edges arranged as a square. Furthermore, to compare with ref.~\cite{Fleury:2016ykk}, let three of the four $\tfrac{1}{2}$-BPS operators be placed at distinguished points,%
\footnote{%
In what follows, all squared distances involving point 4 will be scaled away.
}
\begin{equation}
\label{eq:position01infty}
\alpha_1 = \bar \alpha_1 = z_1 = \bar z_1 = 0,
\qquad
\alpha_3 = \bar \alpha_3 = z_3 = \bar z_3 = 1,
\qquad
\alpha_4 = \bar \alpha_4 = z_4 = \bar z_4 = \infty,
\end{equation}
while the remaining operator is at a generic point
\begin{equation}
\label{eq:positionzzbar}
\alpha_2 = \alpha, \quad
\bar \alpha_2 = \bar \alpha, \qquad
z_2 = z, \quad
\bar z_2 = \bar z\,.
\end{equation}
We therefore have
\begin{equation}
\label{eq:distancestocrossratios}
\begin{aligned}
x_{12}^2 =&\, z \bar z\,, &\qquad
 x_{13}^2 =&\, 1\,, &\qquad
 x_{23}^2 =&\, (1-z)(1-\bar z)\,,\\
y_{12}^2 =&\, \alpha \bar \alpha\,,&\qquad
y_{13}^2 =&\, 1\,,&\qquad
 y_{23}^2 =&\, (1-\alpha)(1-\bar \alpha)\,.
\end{aligned}
\end{equation}
We can rewrite our results for $\tilde{F}_{ij;kl}$ as
\begin{eqnarray}
\tilde F_{12;34} & = & z \bar z \left[ \frac{1}{\alpha} + \frac{1}{\bar \alpha} - \frac{1}{z} - \frac{1}{\bar z} \right] \, , \nonumber \\
\tilde F_{13;24} & = & \alpha +\bar \alpha - z - \bar z \, , \\
\tilde F_{23;14} & = & - (1-z)(1-\bar z) \left[ \frac{1}{1-\alpha} + \frac{1}{1-\bar \alpha} - \frac{1}{1-z} - \frac{1}{1 - \bar z} \right] \, . \nonumber
\end{eqnarray}
In a square with consecutive corners 1243 there can be two planar Yang-Mills exchanges: one from edges 12 to 34, \textit{i.e.}\ $\tilde F_{12;34}$, and the other between edges 13 to 23, \textit{i.e.}\ $\tilde F_{13;24}$. Their sum yields
\begin{equation}
\tilde F_{12;34} + \tilde F_{13;24} = - 2 (z + \bar z) + \left(\frac{1}{\alpha} + \frac{1}{\bar \alpha}\right) (\alpha \bar \alpha + z \bar z)
\end{equation}
which is exactly the rational pre-factor of the box integral in formula (53) of ref.~\cite{Fleury:2016ykk}.
Note that such a pre-factor is the contribution of \emph{twice} the exchange of a mirror magnon. This is the full integrability result, as mirror magnons can be exchanged on the ``front'' and on the ``back'' of the square; the front and back contributions are identical. Let us also remark that the contributions due to the exchange of a mirror magnon over a length-zero diagonal (say, 14) or over the anti-diagonal (23) are also identical. This is readily seen as the cross ratios are invariant under  the simultaneous exchange $1 \leftrightarrow 2$ and $3 \leftrightarrow 4$.
The result is graphically displayed in figure~\ref{fig:hexagonFTmatch}.

We have so far considered the case of the empty box. The computation for a diagram five non-vanishing bridge-lengths---a square with a diagonal---follows the same lines and gives exactly the same result in field theory. This might be puzzling at first, as in the integrability picture we have \emph{a single} mirror magnon exchange---on the face of the square which does not contain a diagonal. However, recall that we have a factor of two due to the fact that there are two inequivalent ways to embed such a ``chiral'' graph in the tessellation of topology~(33), \textit{cf.}\ figure~\ref{fig:fourtessellations}. Hence, also for this case we find perfect agreement.

Finally, it is interesting to turn this picture around and see how a mirror-magnon exchange can be interpreted in terms of Feynman graphs. Let us denote a mirror exchange across the zero-length edge $ij$ as $I_{ij}$. We have seen that
\begin{equation}
I_{23} = \frac{1}{2}\big(\tilde F_{12;34} + \tilde F_{13;24} \big),
\qquad
I_{13} = \frac{1}{2}\big(\tilde F_{12;34} + \tilde F_{23;14} \big),
\qquad
I_{12} = \frac{1}{2}\big(\tilde F_{13;24} + \tilde F_{23;14} \big).
\end{equation}
This system can be inverted to give
\begin{equation}
\tilde F_{12;34} = I_{13} + I_{23} - I_{12} \, .
\end{equation}
Notice that for the edge-reducible topologies, which we have excluded on the grounds of colour scaling, it would not be easy to propose such a matching.

In conclusion \emph{integrability reproduces the field-theory structure graph by graph} for one-loop BPS four-point functions.  We expect the same arguments to apply in the case of $sl(2)$ excitations. As for more general excitations, our arguments may need to be adapted; in particular, as we remarked transverse excitations would be harder to study in the $\cN=2$ formalism, since they are in the YM multiplet, rather than in the matter ones; still, it should be possible to analyse them too, at least at one~loop. Finally, it would very interesting to see whether and how this picture can be extended to higher loops and higher-point functions.

\section{Conclusions and outlook}
\label{sec:conclusions}
One of the main outcomes of our investigation was the realisation that the hexagon prescription used so far in the computation of correlation functions is incomplete. While this did not play a role for (non-extremal planar) three-point functions, it becomes an unavoidable issue for four-point functions and non-planar correlators. We proposed to amend the hexagon prescription by dressing diagrams by $SU(N)$ colour factors. We tested this idea extensively, and found perfect agreement for tree-level field theory; moreover, this correctly explains the empirical rule of ref.~\cite{Fleury:2016ykk} for accounting for wrapping interactions at one loop---though things look more non-trivial at two loops, which is so far unexplored territory.

\begin{figure}[t]
\begin{center}
\includegraphics[width=\linewidth]{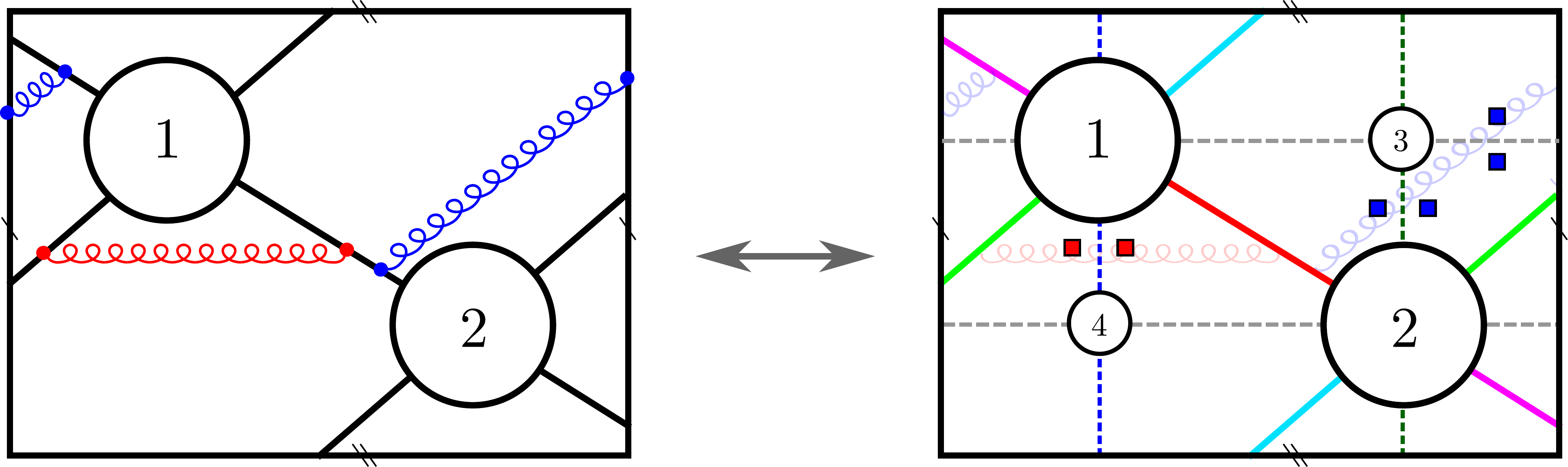}
\caption{We represent the torus as a square, and consider a two-point function in the tessellation of section~\ref{sec:torusnorm}. By the intuition developed in section~\ref{sec:wrapping}, attaching a YM line to a strand of propagators should have an interpretation in terms of virtual magnons. For the usual planar wrapping interactions, this appears straightforward in the spirit of figure~\ref{fig:hexagonFTmatch}. In the case of exchanges that go around the torus' cycles---in field theory, a YM line passing through the sphere---we expect multi-magnon wrapping to appear. This is not surprising, as all dotted lines in the figure are zero-length edges.
}
\label{fig:wrappingtorus}
\end{center}
\end{figure}

We also proposed to employ the hexagon approach to compute next-to-leading order corrections in the $1/N$ expansion for correlation functions, and we tested this idea for tree-level two-point functions finding perfect agreement between field theory and our hexagon-based construction; once again, colour-dressing was crucial. It is a long-standing question whether integrability of $\cN=4$ SYM can be extended beyond the leading order in the large-$N$ expansion. Expecting integrability at finite~$N$ may indeed be far too optimistic; yet there is some hope to systematically build over the large-$N$ integrability to incorporate sub-leading terms. There are two facets to this issue: finding non-planar corrections to conformal eigenstates, and computing correlators involving multi-trace operators and higher-genus worldsheet.
Our construction shows that the hexagon formalism can, in principle, be used for the latter part of this problem.
It would be important to further explore these ideas, both going towards higher genus and incorporating wrapping corrections in the formalism. Both tasks are in principle straightforward, though technically involved: in the former case, we would need very many hexagons to tessellate an high-genus surface; in the latter, due to the insertion of two identity operators to regularise the tessellation, we have many mirror edges of null width. This would lead to a proliferation of wrapping interactions already at one loop, as it is in a sense expected from the field theory intuition for wrapping interactions that we developed in the $\cN=2$ formalism; in figure~\ref{fig:wrappingtorus} we sketch a possible wrapping interaction on the torus.
On top of this, the remaining issue of determining the conformal eigenstates is of crucial importance, and it would be interesting to see if integrability can help there too---at least in a $1/N$~expansion.

Exploring non-planarity remains one of the outstanding challenges for the integrability program. The hexagon might prove instrumental for tackling this problem and we expect exciting developments in the near future.\footnote{We are grateful to P.~Vieira for informing us about an upcoming work in this direction \cite{pedro_paper}.}

\section*{Acknowledgements}

We are grateful to Gleb Arutyunov, Niklas Beisert, Marius de Leeuw, Jan Plefka and Emery Sokatchev for useful and interesting related discussions. We thank Till Bargheer, Jo\~ao Caetano, Thiago Fleury, Shota Komatsu and Pedro Vieira for informing us about their upcoming paper.
B.~Eden is supported by the DFG ``eigene Stelle'' Ed 78/4-3. D.~le Plat acknowledges support by the Stiftung der Deutschen Wirtschaft. A.~Sfondrini acknowledges support by the ETH ``Career Seed Grant'' n.~0-20313-17. Y.~Jiang and A.~Sfondrini acknowledge partial support from the NCCR SwissMAP, funded by the Swiss National Science Foundation.

\appendix

\section{Hexagon form factor at tree level}
\label{app:tree}

The hexagon form factor~\cite{Basso:2015zoa} consists of a matrix part and a scalar ``dressing'' factor. Physical magnons can sit on three edges, and accordingly we write
\begin{equation}
\mathfrak{h}_{123}(\alpha_1, \alpha_2, \alpha_3)\,,
\end{equation}
to indicate a hexagon with edges $1,2,3$ populated by excitations $\alpha_1,\alpha_2,\alpha_3$, respectively. The hexagon is invariant under cyclic shifts of the three sets of physical magnons. The empty hexagon is equal to 1, for one magnon we find
\begin{equation}
\mh(\{Y\}, \emptyset, \emptyset) \, = \, - \mh(\{\bar Y\}, \emptyset, \emptyset) \, = \, 1 \, , \qquad  \mh(\{X\}, \emptyset, \emptyset) \, = \, \mh(\{\bar X\}, \emptyset, \emptyset) \, = \, 0 \, ,
\end{equation}
where we suppressed the subscript index $123$, as we will do in the remainder of this appendix.

Non-trivial dynamics can only arise when there are two or more magnons on the same hexagon. Following ref.~\cite{Basso:2015zoa} (see also ref.~\cite{Arutyunov:2006yd}) we discuss the scattering in the string frame in order to better handle crossing transformations. The outcome is finally converted to the spin chain frame.
The formulae below were used for the evaluation of the amplitudes $\cA^{(ij)}$ and $\cA^8$ described in the main text. For the two-excitation BMN operators we can impose the level-matching conditions $u_2 = - u_1, \, u_4 = - u_3$ from the start; this will simplify our formulae. Furthermore, we will omit the form factors that can be found by substitution such as $u_1\leftrightarrow u_2$.
\begin{align}
\mh(\{X_1\}, \emptyset, \{\bar X_3\}) =&\quad \frac{i}{u_1 - u_3}\,,\\
\mh(\{\bar X_1\}, \emptyset, \{X_3\})  =&\quad   \frac{i}{u_1 - u_3}\,, \\
\mh(\{X_1, X_2\}, \emptyset, \{\bar X_3, \bar X_4\})=&\quad\frac{u_1 u_3 (1 + 2 \, u_1^2 + 2 \, u_3^2)}{(u_1 + \frac{i}{2}) (u_3 + \frac{i}{2}) (u_1 - u_3)^2 (u_1 + u_3)^2 }\,,\\
\displaybreak[2]
\mh(\{\bar X_1, \bar X_2\}, \emptyset, \{X_3, X_4\}) =&\quad  \frac{u_1 u_3 (1 + 2 \, u_1^2 + 2 \, u_3^2)}{(u_1 + \frac{i}{2}) (u_3 + \frac{i}{2}) (u_1 - u_3)^2 (u_1 + u_3)^2 }  \,;\\
\mh(\{\bar Y_1\}, \emptyset, \{Y_3\})=&\quad  -1\,,\\
 \mh(\{Y_1\}, \emptyset, \{\bar Y_3\}) =&\quad  -1 \, ,\\
\mh(\{\bar Y_1, \bar Y_2\}, \emptyset, \emptyset)=&\quad  \frac{u_1}{u_1 + \frac{i}{2}} \, ,\\
\mh(\{Y_1, Y_2\}, \emptyset, \emptyset) =&\quad \frac{u_1}{u_1 + \frac{i}{1}} \, ,\\
\displaybreak[1]
\mh(\{\bar Y_1, \bar Y_2\}, \emptyset, \{Y_3\})=&\quad \frac{u_1}{u_1 + \frac{i}{2}} \, , \\
-\mh(\{Y_1, Y_2\}, \emptyset, \{\bar Y_3\})    =&\quad \frac{u_1}{u_1 + \frac{i}{2}} \, , \\
\displaybreak[1]
-\mh(\{\bar Y_1\}, \emptyset, \{Y_3, Y_4\})      =&\quad \frac{u_3}{u_3 + \frac{i}{2}} \, , \\
 \mh(\{Y_1\}, \emptyset, \{\bar Y_3, \bar Y_4\}) =&\quad \frac{u_3}{u_3 + \frac{i}{2}} \, , \\
\displaybreak[1]
\mh(\{\bar Y_1, \bar Y_2\}, \emptyset, \{Y_3, Y_4\}) =&\quad \frac{u_1 u_3}{(u_1 + \frac{i}{2})(u_3 + \frac{i}{2})} \, , \\
\mh(\{Y_1, Y_2\}, \emptyset, \{\bar Y_3, \bar Y_4\}) =&\quad  \frac{u_1 u_3}{(u_1 + \frac{i}{2})(u_3 + \frac{i}{2})} \, , \\
\displaybreak[1]
\mh(\{\bar Y_1\}, \emptyset, \{\bar Y_3\}) =&\quad \frac{u_1 - u_3 + i}{u_1 - u_3} \, , \\
\mh(\{Y_1\}, \emptyset, \{Y_3\}) 		  =&\quad \frac{u_1 - u_3 + i}{u_1 - u_3} \, , \\
\displaybreak[1]
-\mh(\{\bar Y_1, \bar Y_2\}, \emptyset, \{\bar Y_3\})=&\quad \frac{u_1 (u_1 - u_3 + i) (u_1 + u_3 - i)}{(u_1 + \frac{i}{2}) (u_1 - u_3) (u_1 + u_3)} \, , \\
 \mh(\{Y_1, Y_2\}, \emptyset, \{Y_3\})               =&\quad \frac{u_1 (u_1 - u_3 + i) (u_1 + u_3 - i)}{(u_1 + \frac{i}{2}) (u_1 - u_3) (u_1 + u_3)} \, , \\
\displaybreak[1]
-\mh(\{\bar Y_1\}, \emptyset, \{\bar Y_3, \bar Y_4\}) =&\quad \frac{u_3 (u_1 - u_3 + i)(u_1 + u_3 + i)}{(u_3 + \frac{i}{2})(u_1 - u_3) (u_1 + u_3)} \, , \\
 \mh(\{Y_1\}, \emptyset, \{Y_3, Y_4\})                =&\quad \frac{u_3 (u_1 - u_3 + i)(u_1 + u_3 + i)}{(u_3 + \frac{i}{2})(u_1 - u_3) (u_1 + u_3)} \, , \\
\displaybreak[0]
\mh(\{\bar Y_1, \bar Y_2\}, \emptyset, \{\bar Y_3, \bar Y_4\}) =&\quad \frac{u_1 u_3 \big[(u_1 - u_3)^2+1\big]\big[(u_1 + u_3)^2+1\big]}{(u_1 + \frac{i}{2}) (u_3 + \frac{i}{2}) (u_1 - u_3)^2 (u_1 + u_3)^2 } \, ,\\
\mh(\{Y_1, Y_2\}, \emptyset, \{Y_3, Y_4\})                     =&\quad \frac{u_1 u_3 \big[(u_1 - u_3)^2+1\big]\big[(u_1 + u_3)^2+1\big]}{(u_1 + \frac{i}{2}) (u_3 + \frac{i}{2}) (u_1 - u_3)^2 (u_1 + u_3)^2 } \, .
\end{align}

\section{More on spacetime dressing}
\label{app:position}
In this appendix we match the spacetime dressing of ref.~\cite{Eden:2016xvg} reviewed in section~\ref{sec:positiondressing} with the one of ref.~\cite{Fleury:2016ykk}.
In that reference the authors derive a space-time dependent ``twist'' due to moving a magnon of flavour~$\chi$ from one hexagon to another via the mirror edge with bridge length $\ell$. They derive such twists from the study of mirror magnons; they are given by $e^{ip\ell}\,\mathcal{W}_\chi$ where
\begin{align}
\mathcal{W}_\chi=e^{-E_\chi\log|z|+J_\chi\varphi}e^{i\text{L}_\chi\phi+i\text{R}_\chi\theta}.
\end{align}
Here $E, J, \text{L}, \text{R}$ are the $U(1)$ charges of the magnons, and the various ``chemical potentials'' are functions of the conformal cross-ratios corresponding to such charges. Moreover, the overall result for a given partition is scaled by a factor~$f_\chi$ which depends on which hexagon we choose as a starting point for distributing the magnons.
In order to make contact with that picture, let us take operators $1,3$ and $4$ to be at positions $0,1,\infty$, see eq.~\eqref{eq:position01infty}, while operator 2 is at position $z,\bar{z}$ in Minkowski space and $\alpha,\bar{\alpha}$ in R-symmetry space, see eq.~\eqref{eq:positionzzbar}. By matching the conventions in this way, our results can be readily identified with those of ref.~\cite{Fleury:2016ykk}.
To make contact with the notation of section~\ref{sec:positiondressing}, $z$ and $\bar{z}$ are related to the holomorphic and anti-holomorphic part of the distance, \textit{e.g.}\ $x_{21}^+=z$, $x_{21}^-=\bar{z}$, see also eq.~\eqref{eq:distancestocrossratios}. Similar formulae hold for the R-symmetry cross ratios $\alpha,\bar{\alpha}$ and $y^\pm_{21}$, and so on; by inserting the Drukker-Plefka kinematics in these formulae we find
\begin{equation}
z=\bar{z}=\alpha=\bar{\alpha}=a\,,
\end{equation}
when the four operators are super-translated like in eq.~\eqref{eq:drukkerplefka} with $a_1=0$, $a_2=a$, $a_3=1$ and~$a_4=\infty$.

\begin{figure}[t]
\begin{center}
\includegraphics[scale=0.3]{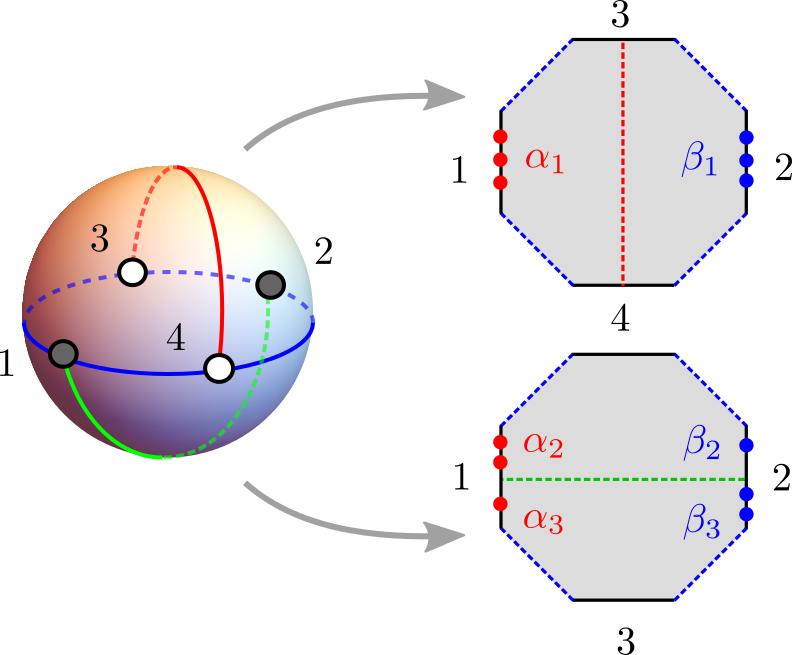}
\caption{One possible tessellation of the four-point function. The vertical black arrows denote the direction in which we move the excitations. Originally all the red excitations are in partition $\alpha_1$, and then we move them downwards by inserting suitable twists in front of the hexagon form factors; similarly for the blue excitations on partitions~$\beta_i$.}
\label{fig:tess_AI}
\end{center}
\end{figure}

In what follows we will consider the case of $sl(2)$ excitations, on which ref.~\cite{Fleury:2016ykk} mostly focuses. It is completely straightforward to repeat the arguments for scalar excitations. Let us consider the set-up of figure~\ref{fig:tess_AI}, for instance, focussing for the time being only on the red magnons in partitions $\alpha_i$. We start from considering $\alpha_1$ magnons on hexagon $\mathfrak{h}_{134}$. Then the ``dressed'' form factor reads
\begin{equation}
f^{(134);\alpha_1}_{D}\, \mathfrak{h}_{134}(\alpha_1)=
\left(\frac{x^{+}_{34}}{x^{+}_{13}x^{+}_{14}}\right)^{|\alpha_1|}
\left(\frac{x^{+}_{34}}{x^{+}_{13}x^{+}_{14}}\frac{x^{-}_{34}}{x^{-}_{13}x^{-}_{14}}\right)^{E(\alpha_1)/2}\, \mathfrak{h}_{134}(\alpha_1),
\end{equation}
where $E(\alpha)=\sum_{j\in\alpha_1}\gamma(u_j)$ is a sum of magnon energies (anomalous dimensions) and we have used the explicit form of $f_\chi$ for an $sl(2)$ excitation. In what follows, we will drop the ``anomalous'' term, $E(\alpha_i)\to0$, as it vanishes at tree level and can anyway be easily reconstructed from the indices of the ``classical'' piece $|\alpha_i|$.
Let us consider the next hexagon. The magnons in $\alpha_2$, that have travelled across edge $14$ down to hexagon $\mathfrak{h}_{142}$, have picked up the usual factor of $(-1)^{|\alpha_2|}e^{ip_2\ell_{14}}$ (where $p_2$ is the total momentum of the magnons in $\alpha_2$) as well as a twist $\mathcal{W}^{(14)}$. They contribute to the partition as
\begin{equation}
\begin{aligned}
& f^{(134);\alpha_2}_{D} (-1)^{|\alpha_2|}e^{ip_2\ell_{14}} \mathcal{W}_{D}^{(14);\alpha_2} \mathfrak{h}_{142}(\alpha_2)\\
&\hspace*{4cm} =
\left(\frac{x^{+}_{34}}{x^{+}_{13}x^{+}_{14}}\right)^{|\alpha_2|}
\left(-\frac{x^{+}_{13}x^{+}_{24}}{x^{+}_{12}x^{+}_{34}}\right)^{|\alpha_2|}e^{ip_2\ell_{14}}
\mathfrak{h}_{142}(\alpha_2)\,.
\end{aligned}
\end{equation}
Notice that the twist can be expressed directly in terms of the cross-ratios, and in particular $\mathcal{W}_{D}^{(14);\alpha_2}=z^{-|\alpha_2|}$.
Finally, the magnons $\alpha_3$ that made all the way down to hexagon $\mathfrak{h}_{123}$ have crossed both edges $14$ and $12$. Accordingly, they picked up a factor of $\mathcal{W}^{(14)}\mathcal{W}^{(12)}$, yielding a contribution of
\begin{equation}
\begin{aligned}
&f^{(134);\alpha_3}_{D} e^{ip_3(\ell_{14}+\ell_{12})} \mathcal{W}_{D}^{(14);\alpha_3}\mathcal{W}_{D}^{(12);\alpha_3}
\mathfrak{h}_{123}(\alpha_3)\\
&\qquad\qquad\qquad\qquad
=\left(\frac{x^{+}_{34}}{x^{+}_{13}x^{+}_{14}}\right)^{|\alpha_3|}
\left(\frac{x^{+}_{13}x^{+}_{24}}{x^{+}_{12}x^{+}_{34}} \frac{x^{+}_{14}x^{+}_{23}}{x^{+}_{13}x^{+}_{24}}\right)^{|\alpha_3|}
e^{ip_3(\ell_{14}+\ell_{12})}
\mathfrak{h}_{123}(\alpha_3)\,.
\end{aligned}
\end{equation}
In terms of the cross-ratios, $\mathcal{W}_{D}^{(12);\alpha_3}=(1-z)^{|\alpha_3|}$.
Indeed by multiplying various pre-factors and taking care of the minus sings, we find that the sum of these three contributions is
\begin{equation}
(v^{-}_{1;34})^{|\alpha_1|} \mathfrak{h}_{134}(\alpha_1)+
e^{ip_2\ell_{14}} (v^{-}_{1;42})^{|\alpha_2|} \mathfrak{h}_{134}(\alpha_2)+
e^{ip_3(\ell_{14}+\ell_{12})} (v^{-}_{1;23})^{|\alpha_3|}\mathfrak{h}_{123}(\alpha_3)\,.
\end{equation}
We can restore the ``anomalous'' part simply by multiplying each $(v^{-}_{i;jk})^{|\alpha_l|}$ by an additional $(v^{+}_{i;jk}v^{-}_{i;jk})^{E(\alpha_l)/2}$. Noticing that in all three hexagons the excitations sit in the first edge, \textit{i.e.}\ $\mathfrak{h}_{134}(\alpha_1)=\mathfrak{h}_{134}(\alpha_1,\emptyset,\emptyset)$, we find perfect matching with equation~\eqref{eq:hexagoncorr}, and we can recast the contribution of this partition simply as
\begin{equation}
\widehat{\mathfrak{h}}_{134}(\alpha_1)+
e^{ip_2\ell_{14}}\,\widehat{\mathfrak{h}}_{134}(\alpha_2)+
e^{ip_3(\ell_{14}+\ell_{12})} \,\widehat{\mathfrak{h}}_{123}(\alpha_3)\,.
\end{equation}

Now,  including the blue excitations in partitions $\beta_{1},\dots\beta_3$ would require a similar exercise. A new feature is that some hexagons will have excitations on more than one physical edge. It is interesting to check our construction in this more general case too. Let us hence consider hexagon $\mathfrak{h}_{142}$ in presence of both sets of excitations. Following the above logic, we obtain
\begin{equation}
\begin{aligned}
&f^{(134);\alpha_2}_{D} (-1)^{|\alpha_2|}e^{ip_2\ell_{14}} \mathcal{W}_{D}^{(14);\alpha_2}\,
f^{(243);\beta_2}_{D} (-1)^{|\beta_2|}e^{iq_2\ell_{42}} \mathcal{W}_{D}^{(42);\beta_2}\,
 \mathfrak{h}_{142}(\alpha_2,\emptyset,\beta_2)\\
&
\qquad=(v_{1;42}^{-})^{|\alpha_2|}e^{ip_2\ell_{14}}\,
\left(\frac{x^{+}_{43}}{x^{+}_{23}x^{+}_{24}}\right)^{|\beta_2|}
\left(-\frac{x^{+}_{23}x^{+}_{41}}{x^{+}_{21}x^{+}_{43}}\right)^{|\beta_2|}e^{iq_2\ell_{42}}\,
\mathfrak{h}_{142}(\alpha_2,\emptyset,\beta_2)\\
&
\qquad=(v_{1;42}^{-})^{|\alpha_2|}e^{ip_2\ell_{14}}\,
(v_{2;41}^{-})^{|\beta_2|}e^{iq_2\ell_{42}}\,\mathfrak{h}_{142}(\alpha_2,\emptyset,\beta_2)\,,
\end{aligned}
\end{equation}
where once again we have discarded the anomalous part to lighten the notation. Again, this is in perfect agreement with the prescription of eq.~\eqref{eq:hexagoncorr}.  The other hexagons, as well as more general partitions and excitation flavours, can be worked out in a similar manner.

\section{Colour factors}
\label{app:colour}
In this appendix we work out how to evaluate the colour factors $T_{ijkl}$ relevant for the hexagon tessellation of the torus.
We start by recalling the contraction rules of $SU(N)$ generators in the adjoint representation,%
\footnote{We have suppressed a factor of 1/2 on the r.h.s.\ of the two contraction rules, and we will not indicate it in the rest of the formulae either. It can easily be re-instated by multiplying the final expressions by $1/2^n$, where $n$ is the number of Wick contractions.
\label{note:factorhalf}
}
\begin{equation}
\begin{aligned}
\text{Tr}\big[A \, T_k\big] \, \text{Tr}\big[B \, T_k\big] = \text{Tr}\big[A B\big] - \frac{1}{N} \text{Tr}\big[A\big] \text{Tr}\big[B\big] \, ,\\
\text{Tr}\big[A \, T_k \, B \, T_k \, C\big] = \text{Tr}\big[B\big] \, \text{Tr}\big[A C\big] - \frac{1}{N} \text{Tr}\big[A B C\big],
\end{aligned}
\label{eq:suNWick}
\end{equation}
where $A,B,C$ are any sequences of colour generators, \textit{i.e.}\ $A=T_{i_1}\cdots T_{i_a}$, and so on.

Let $\Af$ denote a sequence of colour indices, and $\Ab$ the same sequence reversed; $\Bf$ and $\Bb$ are similar sequences, and $\Af, \Bf$ have no element in common. We denote by a $\Af{}'$ the $\Af$ sequence with one element at the end omitted. By the contraction rules (\ref{eq:suNWick}) we have
\begin{equation}
a_j \equiv \text{Tr}\big[\Af\, \Ab\big] = \text{Tr}\big[\Af{}'\, T_k \, T_k\, \Ab{}'\big] = \text{Tr}\big[1\big]\, \text{Tr}\big[\Af{}'\, \Ab{}'\big] - \frac{1}{N} \text{Tr}[\Af{}'\, \Ab{}'] = C \, a_{j-1} \, ,
\end{equation}
where $C$ is the quadratic Casimir of $su(N)$ in the adjoint representation, up to the factor of $1/2$ of footnote~\ref{note:factorhalf},
\begin{equation}
C = \frac{N^2-1}{N}\,.
\end{equation}
Clearly, $a_0 = N$ so that iteration of the last equation implies
\begin{equation}
a_j = C^j \, N \, .
\end{equation}
Further, for $k > 0$ consider
\begin{equation}
\begin{aligned}
b_j &\equiv \text{Tr}\big[\Af\big]\, \text{Tr}\big[\Ab\big]
= \text{Tr}\big[\Af{}'\, T_k\big] \text{Tr}\big[T_k\, \Ab{}'\big]\\
&= \text{Tr}\big[\Af{}'\, \Ab{}'\big] - \frac{1}{N} \text{Tr}\big[\Af{}'\big]\text{Tr}\big[\Ab{}'\big]
= a_{j-1} - \frac{1}{N} b_{j-1}\,,
\end{aligned}
\end{equation}
which can also be iterated with the boundary condition $b_1 = 0$. We find
\begin{equation}
b_j = C \left( C^{j-1} - (-N)^{1-j} \right) \, .
\end{equation}
Next, for $j > 0$ consider
\begin{equation}
\begin{aligned}
c_{i,j} &\equiv\text{Tr}\big[\Bf\big]\, \text{Tr}\big[\Af\, \Ab\, \Bb\big]
= \text{Tr}\big[T_k \Bf{}'\big]\, \text{Tr}\big[\Af\, \Ab\, \Bb{}'\, T_k\big]\\
&= \text{Tr}\big[\Af\, \Ab\, \Bb{}'\, \Bf{}'\big] - \frac{1}{N} \text{Tr}\big[\Bf'\big]\, \text{Tr}\big[\Af\, \Ab\, \Bb'\big] = a_{i+j-1} - \frac{1}{N} c_{i,j-1} \, .
\end{aligned}
\end{equation}
Iterating with $c_{i,1} = 0$ we obtain
\begin{equation}
c_{i,j} = C^{i+1} \left( C^{j-1} - (-N)^{1-j} \right) \, .
\end{equation}
Finally, for $i > 0$
\begin{equation}
\begin{aligned}
d_{i,j} &\equiv \text{Tr}\big[\Af\, \Bf\, \Ab\, \Bb\big]
 = \text{Tr}\big[\Af{}'\, T_k\, \Bf\, T_k\, \Ab{}' \Bb\big]\\
& = \text{Tr}\big[\Bf\big]\, \text{Tr}\Big[\Af{}'\, \Ab{}'\, \Bb\big] - \frac{1}{N} d_{i-1,j} \, .
\end{aligned}
\end{equation}
Repeated application with the boundary condition $d_{0,j} = C^j \, N$ yields
\begin{equation}
d_{i,j} = \frac{C^{i+j}}{N} + \left(\frac{-1}{N}\right)^i C^{j+1} + \left( \frac{-1}{N} \right)^j C^{i+1} - \left( \frac{-1}{N} \right)^{i+j} C \, .
\end{equation}

After these preparations, let us evaluate $T_{ij10}$ for $i,j > 0$.
\begin{equation}
\begin{aligned}
T_{ij10} & =  \text{Tr}\big[\Af\, \Bf\, T_k\big]\, \text{Tr}\big[\Ab\, \Bb\, T_k\big]\\
& = \text{Tr}\big[\Af\, \Bf\, \Ab\, \Bb\big] - \frac{1}{N} \text{Tr}\big[\Af\, \Bf\big]\, \text{Tr}\big[\Ab\, \Bb\big]\\
& = d_{i,j} - \frac{1}{N} b_{i+j}  =  \left(\frac{-1}{N}\right)^i C^{j+1} + \left( \frac{-1}{N} \right)^j C^{i+1} - 2 \left( \frac{-1}{N} \right)^{i+j} C \, .
\end{aligned}
\label{eq:tij1}
\end{equation}
This result comes about because the term $C^{i+j}/N$ cancels between $d_{i,j}$ and $b_{i+j}/N$.
When $i,j > 1$ and for $N\gg1$, that term is actually leading and goes as $C^{i+j}/N = N^{i+j-1} + \ldots$. Recall that $T_{ij10}$ comes from a graph with $i+j+1$ propagators, which at leading order should go like $N^{i+j+1}$ on the sphere and like $N^{i+j-1}$ on the torus. Hence, a leading-order torus contribution cancels. Moreover, the remaining terms are all subleading: since
\begin{equation}
i,j, > 1 \quad \Rightarrow \quad i+1-j < i < i + j -1 \, , \qquad j+1 - i < j < i + j - 1\,,
\end{equation}
the first and second term in the second line of (\ref{eq:tij1}) are suppressed by at least $1/N^4$ w.r.t. to the sphere. Hence we should not consider~$T_{ij10}$ for $i,j > 1$.

Finally, assume $j=1$:
\begin{equation}
T_{i110} = \left(\frac{-1}{N}\right)^i C^{2} - \frac{1}{N} C^{i+1} - 2 \left( \frac{-1}{N} \right)^{i+1}
\end{equation}
and so
\begin{equation}
T_{1110} = -2 N + \ldots \, , \qquad\text{and}\quad T_{i110} = - N^i + \ldots \ \text{for} \ i > 1\,.
\end{equation}


\makeatletter \@ifundefined{Sphere}{\newcommand{\Sphere}{\text{S}}}{}
  \@ifundefined{AdS}{\newcommand{\AdS}{\text{AdS}}}{}
  \@ifundefined{CFT}{\newcommand{\CFT}{\text{CFT}}}{}
  \@ifundefined{CP}{\newcommand{\CP}{\text{CP}}}{}
  \@ifundefined{Torus}{\newcommand{\Torus}{\text{T}}}{}
  \@ifundefined{superN}{\newcommand{\superN}{\mathcal{N}}}{}
  \@ifundefined{grpOSp}{\newcommand{\grpOSp}{\mathrm{OSp}}}{}
  \@ifundefined{grpPSU}{\newcommand{\grpPSU}{\mathrm{PSU}}}{}
  \@ifundefined{grpSU}{\newcommand{\grpSU}{\mathrm{SU}}}{}
  \@ifundefined{grpU}{\newcommand{\grpU}{\mathrm{U}}}{}
  \@ifundefined{grpD}{\newcommand{\grpD}{\mathrm{D}}}{}
  \@ifundefined{grpSL}{\newcommand{\grpSL}{\mathrm{SL}}}{}
  \@ifundefined{grpSp}{\newcommand{\grpSp}{\mathrm{Sp}}}{}
  \@ifundefined{grpUSp}{\newcommand{\grpUSp}{\mathrm{USp}}}{}
  \@ifundefined{grpSO}{\newcommand{\grpSO}{\mathrm{SO}}}{}
  \@ifundefined{grpO}{\newcommand{\grpO}{\mathrm{O}}}{}
  \@ifundefined{algOSp}{\newcommand{\algOSp}{\mathfrak{osp}}}{}
  \@ifundefined{algPSU}{\newcommand{\algPSU}{\mathfrak{psu}}}{}
  \@ifundefined{algSU}{\newcommand{\algSU}{\mathfrak{su}}}{}
  \@ifundefined{algSp}{\newcommand{\algSp}{\mathfrak{sp}}}{}
  \@ifundefined{algSL}{\newcommand{\algSL}{\mathfrak{sl}}}{}
  \@ifundefined{algGL}{\newcommand{\algGL}{\mathfrak{gl}}}{}
  \@ifundefined{algU}{\newcommand{\algU}{\mathfrak{u}}}{}
  \@ifundefined{algSO}{\newcommand{\algSO}{\mathfrak{so}}}{}
  \@ifundefined{algO}{\newcommand{\algO}{\mathfrak{o}}}{}
  \@ifundefined{Integers}{\newcommand{\Integers}{\mathbb{Z}}}{}
  \@ifundefined{Reals}{\newcommand{\Reals}{\mathbb{R}}}{} \makeatother

\providecommand{\href}[2]{#2}\begingroup\raggedright\endgroup

\end{document}